\documentclass[11pt,a4paper]{article}
\usepackage{jheppub}
\usepackage[latin1]{inputenc}
\usepackage{amsmath}
\usepackage{amsfonts}
\usepackage{dsfont}
\usepackage{float}
\usepackage{amssymb}
\usepackage{graphicx}
\usepackage[yyyymmdd,hhmmss]{datetime}
\newcommand{\N}{\mathbb{N}}
\newcommand{\C}{\mathbb{C}}
\newcommand{\R}{\mathbb{R}}
\newcommand{\Z}{\mathbb{Z}}
\newcommand{\sign}{\text{sign}}
\newcommand{\la}{\langle}
\newcommand{\ra}{\rangle}
\newcommand{\mc}[1]{\mathcal{#1}}
\newcommand{\ket}[1]{|#1\rangle}
\newcommand{\bra}[1]{\langle#1|}
\title{Entanglement and correlations in the continuous multi-scale entanglement renormalization ansatz}
\author[a,b]{Adri\'an Franco-Rubio}
\author[a]{Guifr\'e Vidal}
\affiliation[a]{Perimeter Institute for Theoretical Physics,\\
	Waterloo, Ontario N2L 2Y5, Canada}
\affiliation[b]{Department of Physics and Astronomy, University of Waterloo, Waterloo, ON N2L 3G1, Canada}
\emailAdd{afrancorubio@pitp.ca}
\abstract{We investigate the entanglement structure of the continuous multi-scale entanglement renormalization ansatz (cMERA)  [Haegeman et al., Phys.\hspace{-0.5mm} Rev.\hspace{-0.5mm} Lett., 110, 100402 (2013)] for ground states of quantum field theories (QFTs). The cMERA, proposed as an extension to QFTs of the lattice MERA, is defined directly in the continuum but is nevertheless naturally equipped with a short-distance scale $1/\Lambda$ that acts as a UV regulator for quantum fluctuations. We consider the simplified setting of Gaussian cMERA for free QFTs, where explicit calculations can be performed. For relativistic free massless bosonic and fermionic QFTs in both 1+1 and 2+1 spacetime dimensions, we show that the cMERA state indeed displays no UV divergences in two-point correlation functions or entanglement entropy, in sharp contrast with the exact ground states.}

\begin{document}
	\maketitle

	\section{Introduction}
	
Tensor networks have emerged in recent years as powerful classes of variational states that can be used to numerically simulate quantum many-body systems. An example is the Multiscale Entanglement Renormalization Ansatz (MERA) \cite{Vidal2005,Vidal2006}, which aims at approximating ground states of local Hamiltonians on the lattice. Its success in a wide range of lattice models, including systems with topological order \cite{Aguado2008,Konig2009} or at a quantum critical point \cite{Giovannetti2008,Pfeifer2009,Evenbly2011}, has established MERA as a useful computational tool. In addition, this success has also provided valuable theoretical insights into the structure of generic many-body ground states. 

For instance, MERA suggests that we can regard the ground state of a local Hamiltonian as the result of an \textit{entangling evolution in scale}. This unitary evolution starts from a product (i.e. unentangled) state and introduces correlations/entanglement into the wavefunction scale by scale, beginning with long distances and progressing towards shorter distances. The entangling evolution then stops when it reaches the shortest length scale or UV cutoff available in the lattice, namely the distance between two neighbouring lattice sites. Moreover, by reversing the entangling evolution, and thus flowing now from short distances to long distances, we obtain a modern implementation of the renormalization group (RG) in the Hilbert space of a quantum many-body system, one that suitably disposes of short-range entanglement at each coarse-graining step. Removal of short-range entanglement has indeed been seen to be key to producing an RG flow with the correct structure of fixed points, including unstable RG fixed points for continuous quantum phase transitions, where the method explicitly produces scale invariance \cite{Vidal2005,Vidal2006}. Finally, MERA has also attracted the interest of the high energy physics community, since it has been conjectured to be a realization of the holographic principle. In the context of the AdS/CFT correspondence \cite{Maldacena1997}, MERA was first proposed as a discretization of (a time slice of) AdS \cite{Swingle2012} and, more recently, as a discretization of the kinematic space of AdS \cite{Czech2016}. 

A continuous version of the MERA, known as cMERA, was proposed by Haegeman, Osborne, Verschelde, and Verstraete in \cite{Haegeman2011}. This proposal has the potential to reproduce, directly in the continuum, the success of MERA on the lattice and thus become a powerful non-perturbative approach to interacting quantum field theories (QFTs). To date, however, cMERA is only well-understood for free QFTs, thanks to a collection of explicit constructions provided in \cite{Haegeman2011}. For a free QFT, the entangling evolution in scale is generated by a quadratic operator (the sum of a generator of scale transformations plus the so-called \textit{entangler}), and thus the resulting state is a Gaussian state. The Gaussian cMERA is of limited interest as a variational ansatz (see nevertheless Ref. \cite{Cotler2016}), but already provides a most valuable proof of principle that both the \textit{entangling evolution in scale} representation of ground states and the closely related \textit{entanglement renormalization} (modern implementation of the renormalization group on wave-functions) can be realized in QFTs directly in the continuum. In addition, the Gaussian cMERA is being actively studied as a possible realization of holography \cite{Miyaji2015,Miyaji2016,Mollabashi2014,Nozaki2012,Caputa2017,Molina-Vilaplana2015,Molina-Vilaplana2016,Miyaji2015a,Miyaji2015b,Swingle2012a,Bao2015,Wen2016a,Gan2016,Fliss2016}.

In this paper we investigate in which sense cMERA has a built-in UV cutoff. On the lattice, the UV cutoff of MERA is quite explicit---there simply are no degrees of freedom between lattice sites! The analogous statement for cMERA is a bit more subtle, since in the continuum there are quantum field degrees of freedom at all distances. However, as pointed out in Ref. \cite{Haegeman2011}, in the cMERA state only those field degrees of freedom corresponding to distances larger than some UV cutoff are significantly entangled. This results from the fact that the entangling evolution that produced the cMERA state in the first place started from a product state (where all the field degrees of freedom are unentangled in real space). As the entangling evolution proceeds, the field degrees of freedom at different length scales become entangled, starting from large distances and then progressively descending to shorter distances. However, the entangling evolution stops at some point, corresponding to some length scale $1/\Lambda$, leaving the degrees of freedom at shorter distances unentangled. The resulting cMERA state is to be compared to the ground state of a relativistic QFT, where the field degrees of freedom are entangled at arbitrarily small length scales. The goal of this paper is to explain and illustrate, through explicit calculations of correlations and entanglement entropy, what it means for the cMERA to have this UV cutoff. 

	\begin{figure}
		\centering
		\includegraphics[width=0.9\linewidth]{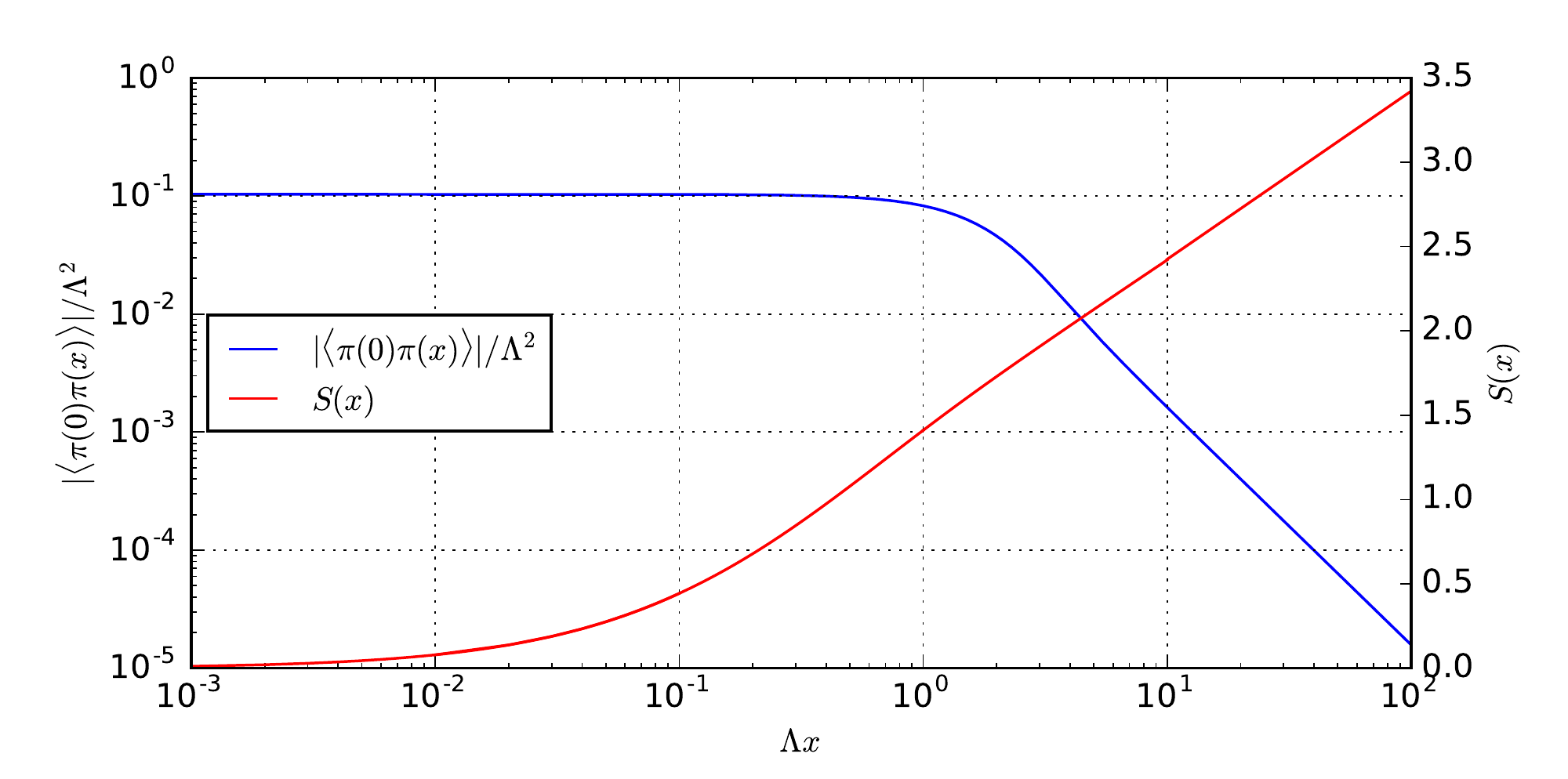}
		\caption{\textit{Left vertical axis.} Plot of the $\langle\pi\pi\rangle$ two-point correlation function for a cMERA with target theory a 1+1 free boson. \textit{Right vertical axis.} Entanglement entropy of an interval vs. length of the interval. for the same cMERA state.}
		\label{intro}
	\end{figure}
	
	As an example that summarizes our findings, let us consider a cMERA $\ket{\Psi^{\Lambda}}$ that has been optimized to approximate the ground state $\ket{\Psi}$ of a conformal field theory (CFT) in 1+1 space-time dimensions. Recall that in the ground state of a 1+1 dimensional CFT, the entanglement entropy of a region of size $x$ is infinite and diverges as
	\begin{equation}
	S(x) = \frac{c}{3} \log \frac{x}{a}
	\end{equation}
as a function of a UV cutoff $a$. Here the UV cutoff $a$ could be the spacing between sites of an auxiliary lattice used to regularize the CFT. As we take $a$ to zero in order to recover the continuum limit, the entropy $S(x)$ of a finite region of size $x$ becomes infinite. On the other hand, the two-point correlation function $C(x)$ of a generic scaling operator with scaling dimension $\Delta > 0$ reads
\begin{equation}
 	C(x) = \frac{1}{x^{2\Delta}}
\end{equation}
and thus diverges at short distances. In contrast, Figure \ref{intro} illustrates the scaling of entanglement entropy and correlations in the cMERA state $\ket{\Psi^{\Lambda}}$ for the the ground state $\ket{\Psi}$ of the 1+1 dimensional free boson CFT, which is the massless limit of the Klein-Gordon QFT, with field $\phi(x)$ and conjugate momentum field $\pi(x)$. As discussed in Section \ref{sec3}, for distances $x$ larger than some length scale $1/\Lambda$, the entanglement entropy and the two-point correlators (exemplified by $\langle \pi(0)\pi(x)\rangle$, where $\pi(x)$ has scaling dimension $\Delta=1$) reproduce the expected logarithmic and power law scaling of the CFT, respectively. However, we find the entanglement entropy $S(x)$ to be finite (as opposed to UV divergent) and it is seen to vanish in the limit of distances $x$ smaller than $1/\Lambda$. Similarly, the two-point correlator $\langle \pi(0)\pi(x)\rangle$ is seen to transition from power-law scaling at large distances to becoming a constant for $x\ll 1/\Lambda$ (with a delta function at $x=0$, not shown in the figure). For both quantities, it is as if we had introduced a lattice with lattice spacing $\approx 1/\Lambda$, although we have not: the cMERA is defined in the continuum.
	
	In what follows we will compute and analyze correlators and entanglement entropy in the Gaussian cMERA approximation to the ground state of bosonic and fermionic free QFTs in 1+1 and 2+1 dimensions, focusing on CFTs for simplicity. Section \ref{sec2} gives some preliminary introduction to the cMERA construction and its interpretation. Section \ref{sec3} builds more explicitly the cMERA for free boson theories in arbitrary dimensions and studies its structure for one and two spatial dimensions. Section \ref{sec4} follows the same scheme for free Dirac fermion theories. Section \ref{sec5} gives a few concluding remarks on the results presented in the paper.

	\section{cMERA preliminaries}
	\label{sec2}
	We begin by briefly reviewing the cMERA, as proposed in Ref. \cite{Haegeman2011}. The cMERA formalism in its full generality applies to generic interacting QFTs. However, it is only well-understood for free QFTs, for which explicit Gaussian cMERA states were described in Ref. \cite{Haegeman2011}. In this paper we focus on Gaussian cMERA states for free CFTs. Thus our analysis is simplified by two facts: the Gaussian character of the cMERA, and the scale invariance of the CFT, which translates into a notion of scale invariance of the cMERA \cite{Hu2017}.

	\subsection{cMERA}
\begin{figure}
			\centering
			\includegraphics[width=0.95\linewidth]{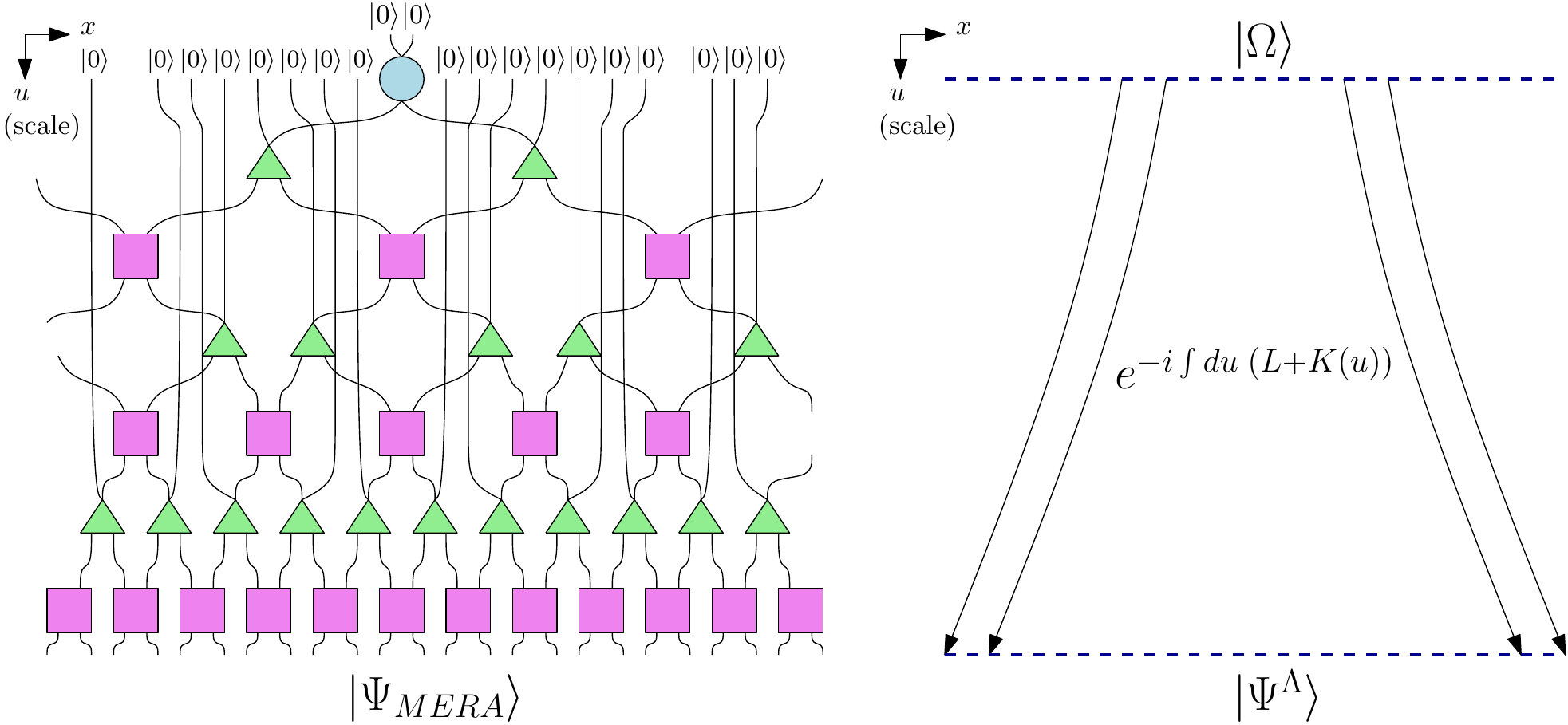}
			\caption{\textit{Left}. A MERA seen as a quantum circuit. Starting from the product state on the top, each layer of unitaries introduces entanglement at shorter and shorter distances. \textit{Right}. A schematic representation of cMERA that emphasizes its similarity with MERA.}
			\label{MERA}
		\end{figure}

	On a lattice made of $N$ sites, MERA can be seen as a quantum circuit that unitarily evolves the product state $\ket{0}^{\otimes N}$, introducing entanglement at different length scales (Figure \ref{MERA}, left). The continuous MERA was introduced as a variational ansatz $\ket{\Psi^\Lambda}$ for the ground state $\ket{\Psi}$ of a quantum field theory, which we will call the \textit{target state}. The cMERA state $\ket{\Psi^{\Lambda}}$  is defined as the result of a continuous unitary evolution in scale $u$ from a product state $\ket{\Omega}$:
	\begin{equation}
	\ket{\Psi^\Lambda}:=U(0,-\infty)\ket{\Omega}\,,\qquad U(u_1,u_2)=\mathcal{P}\text{exp}\left(i \int_{u_1}^{u_2}{du\;(L+K(u))}\right)\,,
	\label{inimera} 
	\end{equation} 
	where $\mathcal{P}\text{exp}$ denotes the path-ordered exponential, and the operators $L$ and $K(u)$ are the generator of scale transformations and the so-called entangler, respectively (Figure \ref{MERA}, right). The scaling operator $L$ depends only on the field content of the QFT (e.g. number of bosonic/fermionic fields and spacetime dimensions) and is independent of the scale parameter $u$. In contrast, the entangler $K(u)$ contains the variational parameters of the cMERA (and thus depends much more on the specific QFT under consideration) and may in general depend also on the scale parameter $u$.

\subsection{Scale invariance}
	
In this paper we will study properties of the cMERA states when the target QFT is a CFT, which is a scale invariant theory. In this case, the entangler $K(u)$ can be chosen to be independent of the scale parameter $u$, substantially simplifying our analysis -- even though the entanglement UV cutoff is similarly present in the case of more general QFTs.

The scale-invariant generator $L+K$ of the entangling evolution in scale is then identified with the generator of scale transformations of the CFT in a very precise sense, which we briefly review. As explained in \cite{Hu2017}, the optimized cMERA $\ket{\Psi^\Lambda}$ can exactly reproduce the spacetime symmetries of the target state $\ket{\Psi}$. However, while a generator $G$ of a symmetry of $\ket{\Psi}$ is typically the integral of a local density, the corresponding symmetry of $\ket{\Psi^{\Lambda}}$ is generated by an operator $G^{\Lambda}$ that is instead the integral of a quasi-local density. In a CFT, the ground state $\ket{\Psi}$ is invariant under global conformal transformations, which include dilations/scale transformations as generated by the dilation operator $D$. Correspondingly, a cMERA $\ket{\Psi^{\Lambda}}$ optimized to approximate the CFT ground state $\ket{\Psi}$ is invariant under a quasi-local realization of global conformal transformations and, in particular, under scale transformations as generated by a quasi-local operator $D^\Lambda$, which can be seen to correspond to $L+K$. Thus $L+K$ can be regarded as the cMERA equivalent of the CFT dilation operator $D$ and, accordingly, the cMERA state $\ket{\Psi^{\Lambda}}$ is scale invariant (that is, by scale transformations generated by $D^{\Lambda} = L+K$).

How can the cMERA state $\ket{\Psi^{\Lambda}}$ be scale invariant and, at the same time, be equipped with an explicit short distance cutoff $1/\Lambda$ in entanglement, as illustrated in Figure \ref{intro}?  As argued in \cite{Hu2017}, this is possible by the combined action of operators $L$ and $K$. The scaling operator $L$ rescales space, which has the effect of shifting the entanglement UV cutoff $1/\Lambda$. However, the entangler $K$ acts as to suitably add or remove short-distance entanglement so as to reset the UV cutoff back at the initial length scale $1/\Lambda$. As a result, scale invariance is compatible with the presence of an explicit length scale.

	\subsection{Gaussian cMERA}

In this work we can restrict our attention to states of a quantum field that are Gaussian states. Indeed, the ground state $\ket{\Psi}$ of a free particle QFT is Gaussian and, as a result, both the initial product state $\ket{\Omega}$ and the cMERA state $\ket{\Psi^{\Lambda}}$ in Eq. \ref{inimera} can be chosen to also be Gaussian.

Recall that a generic Gaussian state $\ket{\Phi}$ is characterized by a complete set of linear constraints, implemented by annihilation operators that are linear combinations of the bosonic/fermionic field operators. When the Gaussian state is invariant under translations, each annihilation operator can be labeled by its momentum $\vec{k}$, and the linear constraints read
	\begin{equation}
	a(\vec k)\ket{\Phi}=0\qquad\forall\vec k\in\R^d.
	\end{equation}
This characterization is particularly useful to study the Gaussian cMERA. The unitary in \eqref{inimera} induces a canonical transformation that maps the linear constraints of $\ket\Omega$ into the linear constraints of $\ket{\Psi^\Lambda}$. For each Gaussian state of interest, we will be able to characterize these constraints in terms of just a real function of the momentum coordinate (which we call $\alpha(\vec k)$ in the bosonic case, and $\theta(\vec k)$ in the fermionic case). We will then see that the linear constraints of the cMERA are just an interpolation between those of the unentangled state $\ket{\Omega}$ for large momenta ($|\vec k|\gg\Lambda$), and those of the target state $\ket{\Psi}$ at small momenta ($|\vec k|\ll\Lambda$), as it was pointed out in \cite{Hu2017} for the 1+1 dimensional free boson theory. This interpolating character already provides us with some intuition about the UV regularization in cMERA states: in $\ket{\Psi^{\Lambda}}$, the large momentum / short distance modes satisfy constraints similar to those of the unentangled product state $\ket{\Omega}$, which is UV finite. 
	
	A well-known, major advantage of working with Gaussian states is that we can efficiently specify them using the above linear constraints/annihilation operators (recall that such an efficient characterization is not available for the ground state of a generic interacting QFT). In addition, the computation of correlations and entanglement entropy from a Gaussian state is similarly well-understood. The reader is invited to check Appendix \ref{apa} for a review of correlations and entanglement entropy in Gaussian states.
	
	We finish this section with a note for the reader: in what comes next we will refer to a cMERA as \textit{optimized} if it succeeds at reproducing the long distance behaviour of the target state, rather than implying that it has gone through an energy minimization procedure.
	\section{Free bosonic QFTs}
	\label{sec3}
	In this section we first review the formalism of cMERA for Klein-Gordon theories in general dimension, drawing mainly from the appendices of \cite{Haegeman2011} and \cite{Hu2017}. We then also review the scaling of correlations and entanglement entropy in the target ground state $\ket{\Psi}$ of the free boson CFT in 1+1 and 2+1 dimensions, and the unentangled product state $\ket{\Omega}$. Finally, we compute and discuss the scaling of correlations and entanglement entropy in the corresponding cMERA state $\ket{\Psi^{\Lambda}}$ in 1+1 and 2+1 dimensions.
	
	 Throughout this section, $\phi(\vec{x})$ denotes a bosonic field operator in $d$ spatial dimensions and $\pi(\vec{x})$ its canonical momentum conjugate field, with $[\phi(\vec{x}), \pi(\vec{y})] = i \delta(\vec{x}-\vec{y})$. Similarly, $\phi(\vec{k})$ and $\pi(\vec{k})$ denote the Fourier components of these field operators, with $[\phi(\vec{k}),\pi(\vec{q})] = i \delta(\vec{k} + \vec{q})$, where
\begin{equation}
\phi(\vec{k}) \equiv \frac{1}{\left(2\pi\right)^{d/2}}\int d^{d}x~ e^{-i\vec{k}\cdot\vec{x}} \phi(\vec{x})\, ,~~~~~~
\pi(\vec{k}) \equiv \frac{1}{\left(2\pi\right)^{d/2}}\int d^{d}x~ e^{-i\vec{k}\cdot\vec{x}} \pi(\vec{x})\,.
\end{equation}

	\subsection{Bosonic Gaussian cMERA framework}
	
	The three bosonic Gaussian states $\ket{\Phi}$ under consideration are determined by linear constraints of the form:
	\begin{equation}
	a(\vec k)\ket{\Phi}=0\qquad a(\vec k)= \sqrt{\dfrac{\alpha(k)}{2}}\phi(\vec k)+i\sqrt{\dfrac{1}{2\alpha(k)}}\pi(\vec k) \qquad\forall \vec k\in\R^d
	\label{form}
	\end{equation}
where the annihilation operator $a(\vec{k})$ is a linear combination of $\phi(\vec{k})$ and $\pi(\vec{k})$ and $k \equiv |\vec{k}|$ is the modulus of the momentum $\vec{k}$. The function $\alpha:\R\to\R$, which we next specify for the target CFT ground state $\ket{\Psi}$, the product state $\ket{\Omega}$, and the cMERA state $\ket{\Psi^{\Lambda}}$, completely determines the Gaussian state under consideration. 

	\subsubsection*{Target state $\ket{\Psi}$}
	We begin by considering the massless Klein-Gordon Hamiltonian in $d$ spatial dimensions:
	\begin{equation}
	H=\dfrac{1}{2}\int{d^dx\;:\left( \pi(\vec x)^2+(\vec\nabla\phi(\vec x))^2\right): }
	\end{equation}
	which upon Fourier transformation allows for diagonalization via creation-annihilation operators 
	\begin{equation}
	H=\dfrac{1}{2}\int{d^dk\;:\left( \pi(\vec k)\pi(-\vec k)+k^2\phi(\vec k)\phi(-\vec k)\right): }=\int{d^dk\;ka^\dagger(\vec k)a(\vec k)}
	\end{equation}
where
	\begin{align}a(\vec k)=\sqrt{\dfrac{k}{2}}\phi(\vec k)+i\sqrt{\dfrac{1}{2k}}\pi(\vec k),\qquad a^\dagger(\vec k)=\sqrt{\dfrac{k}{2}}\phi(-\vec k)-i\sqrt{\dfrac{1}{2k}}\pi(-\vec k).
	\label{creacftddim1}
	\end{align}
Therefore, we can define the target state as a common kernel of the annihilation operators of the form \eqref{form} with 

\begin{equation}
\alpha(k) = k~~~~~~~~~~\mbox{(target CFT ground state $\ket{\Psi}$)}.
\end{equation}

	\subsubsection*{Product state $\ket{\Omega}$}
	Following \cite{Haegeman2011} we consider the Gaussian product state defined by
	\begin{align}\left( \sqrt{\dfrac{\Lambda}{2}}\phi(\vec x)+i\sqrt{\dfrac{1}{2\Lambda}}\pi(\vec x)\right)\ket{\Omega}=0\qquad\forall \vec x\in\R^d
	\end{align}
	with $\Lambda$ constant. That is, the annihilation operators in a Gaussian product state are local in position space. This guarantees that both the connected two-point correlators $\langle\mathcal{O}(x)\mathcal{O}'(y)\rangle-\langle\mathcal{O}(x)\rangle\langle\mathcal{O}'(y)\rangle$ and the entanglement entropy $S(x)$ vanish identically, as expected in a product state. In momentum space the above constraints read
	\begin{align}\left( \sqrt{\dfrac{\Lambda}{2}}\phi(\vec k)+i\sqrt{\dfrac{1}{2\Lambda}}\pi(\vec k)\right)\ket{\Omega}=0\qquad\forall \vec x\in\R^d.
	\end{align}
	That is, 
\begin{equation}
	\alpha(k) = \Lambda~~~~~~~~~~\mbox{(initial product state $\ket{\Omega}$)}
\end{equation}	
for some constant $\Lambda$.

\subsubsection*{cMERA}
	We consider an optimized cMERA for critical bosons based on \cite{Haegeman2011}, which is given by the choice of scaling and entangling operators:
	\begin{align}
	L&=\dfrac{1}{2}\int{d^dk\left[\pi(-\vec k)\left(\vec k\vec \nabla_k+\frac{1}{2}\right)\phi(\vec k)+\text{h.c.}\right]},\nonumber\\ K&=\dfrac{1}{2}\int{d^dk\;g(k)\left[\pi(-\vec k)\phi(\vec k)+\text{h.c.}\right]},
	\end{align}
where $g(k)$ reads
	\begin{equation}
	g(k)=\dfrac{1}{2}e^{-\frac{1}{\sigma}\frac{k^2}{\Lambda^2}}.
	\end{equation}
The constant $\sigma$ in $g(k)$, not present in Ref. \cite{Haegeman2011}, is added here so that the cMERA $\ket{\Psi^{\Lambda}}$ properly matches the long-distance properties of its target state $\ket{\Psi}$. With this particular choice of $g(k)$, the cMERA wavefunction is characterized by the function
	\begin{equation}
	\alpha(k)=\Lambda\exp{\left(\frac{1}{2}\text{Ei}\left(-\frac{1}{\sigma}\frac{k^2}{\Lambda^2} \right) \right)}
	\label{micmera}
	\end{equation}
	where $\text{Ei}$ is the exponential integral function \cite{Hu2017}. For small $k$, this goes as 
	\begin{equation}
	\alpha(k)\sim \sqrt{\dfrac{e^{\gamma}}{\sigma}}k
	\end{equation}
	where $\gamma\approx0.57722$ is the Euler-Mascheroni constant. Hence we choose $\sigma=e^\gamma$ for $\alpha(k)$ to reproduce the behaviour of the target state at small $k$. For $k\to\infty$, $\alpha(k)$ tends to the constant $\Lambda$. Thus the function $\alpha(k)$ for the cMERA satisfies
\begin{equation}
	\alpha(k) = \left\{ 
	\begin{array}{cc}
	k &~~~\mbox{for}~ k \ll \Lambda\\
	\Lambda & ~~~\mbox{for}~k\gg \Lambda
	\end{array}	\right.~~~~~~~~~~\mbox{(optimized cMERA state $\ket{\Psi^{\Lambda}}$)}.
	\end{equation}	
	
	\begin{figure}
\centering
\includegraphics[width=0.5\linewidth]{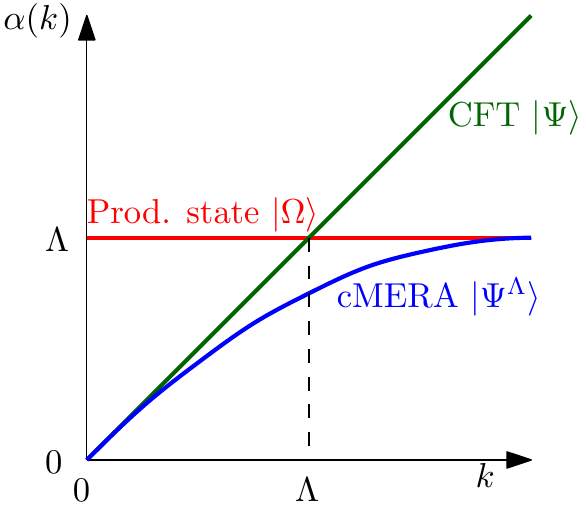}
\caption{Qualitative plot of the characteristic function $\alpha(k)$ for the three states in the cMERA construction.}
\label{alpha}
\end{figure}

	Figure \ref{alpha} shows the qualitative behaviour of $\alpha(k)$ for the three states. The cMERA $\ket{\Psi^{\Lambda}}$ clearly appears as an interpolation between the target CFT ground state $\ket{\Psi}$ at small momentum (approximate linear scaling $\sim  k$ for $k\ll \Lambda$) and the product state $\ket{\Omega}$ at large momentum (approaching the constant $\Lambda$ for $k\gg \Lambda$).
	
	\subsection{Correlations and entropy}
	To probe the short-distance structure of the cMERA state, we will investigate the scaling of two-point correlation functions and entanglement entropy. For comparison purposes we briefly review the scaling of two-point correlations and entanglement entropy in the two states $\ket{\Psi}$ and $\ket{\Omega}$ of which cMERA is an interpolation.

	\subsubsection*{Target state $\ket{\Psi}$}
	 
In the ground state $\ket{\Psi}$ of a free bosonic CFT in $d+1$ spacetime dimensions, the two-point correlation functions involving $\phi(\vec{x})$ and $\pi(\vec{x})$ read
	\begin{equation}
	\la\phi(\vec x)\phi(\vec y)\ra\propto\dfrac{1}{|\vec x-\vec y|^{d-1}}\qquad 	\la\phi(\vec x)\pi(\vec y)\ra=\dfrac{i\delta(\vec x-\vec y)}{2}\qquad 	\la\pi(\vec x)\pi(\vec y)\ra\propto\dfrac{1}{|\vec x-\vec y|^{d+1}}
	\end{equation}  
except in $d=1$, where the first two-point correlation function is instead
	\begin{equation}
	\la\phi(x)\phi(y)\ra=-\dfrac{\log|x-y|}{2\pi}.
	\end{equation}	
In turn, the entanglement entropy of a region of linear size $x$ obeys the area law:
	\begin{equation} \label{arealaw}
	S(x)\sim \left(\frac{x}{a}\right)^{d-1}
	\end{equation}
where $a$ is a UV cutoff, except in $d=1$, where the scaling is logarithmic:
	\begin{equation}
	S(L)\sim\dfrac{c}{3}\log\left(\dfrac{L }{a}\right)
	\label{ctercioslog}
	\end{equation}
	with $c=1$ the central charge of the free boson CFT. Note that if we remove the UV cutoff by taking the limit $a \to 0$, the entanglement entropy in Eqs. \ref{arealaw}-\ref{ctercioslog} diverges. 
	\subsubsection*{Product state $\ket{\Omega}$}
	In the case of the product state $\ket\Omega$ we have
	\begin{equation}
	\la\mc O(\vec x)\mc O'(\vec y)\ra-\la\mc O(\vec x)\ra\la\mc O'(\vec y)\ra = 0.
	\end{equation}
It follows that any connected correlator, as well as the entanglement entropy of any interval, are all zero.
 
	\vspace{0.5cm}
	
	After all these preliminaries, let us implement what we have been presenting for bosonic theories in 1+1 and 2+1 dimensions.
	
	\subsection{Free massless boson in 1+1 dimensions}
The position space two-point functions of the cMERA state given by \eqref{micmera} in 1+1 dimensions are most easily computed by Fourier transforming the corresponding momentum space correlators
\begin{equation}
\la\phi(k)\phi(q)\ra =\dfrac{1}{2\alpha(k)}\delta(k+q),~~ \la\phi(k)\pi(q)\ra = \dfrac{i}{2}\delta(k+q),~~
\la\pi(k)\pi(q)\ra = \dfrac{\alpha(k)}{2}\delta(k+q).
\label{momspa}
\end{equation}
For instance, 
\begin{align}
	\la\phi(x)\phi(y)\ra&=\int{d k\,d q ~\dfrac{e^{i (kx+qy)}}{2\pi} \la\phi(k)\phi(q)\ra }=\int{d k ~\dfrac{e^{i k (x-y)}}{2\pi} \dfrac{1}{2\alpha(k)} }=\nonumber\\&=\dfrac{1}{2\Lambda}\delta(x-y)+\int_{\R\backslash(-\varepsilon\Lambda,\varepsilon\Lambda)}{d k ~\dfrac{e^{i k (x-y)}}{2\pi}\left( \dfrac{1}{2\alpha(k)}-\dfrac{1}{2\Lambda}\right) }=\nonumber\\&=\dfrac{1}{2\Lambda}\delta(x-y)+f_\varepsilon(x-y).
\end{align}
And equally,
		\begin{align}
		\la\phi(x)\pi(y)\ra&=i\dfrac{\delta(x-y)}{2}\\ \la\pi(x)\pi(y)\ra&=\dfrac{\Lambda}{2}\delta(x-y)+\int_{\R\backslash(-\varepsilon\Lambda,\varepsilon\Lambda)}{d k ~\dfrac{e^{i k (x-y)}}{4\pi}\left( \alpha(k)-\Lambda\right)}=\dfrac{\Lambda}{2}\delta(x-y)+g_\varepsilon(x-y).
		\end{align}		
		$f_\varepsilon,g_\varepsilon$ are continuous functions that depend on a parameter $\varepsilon\ll1$. Here $\varepsilon\Lambda$ acts as an IR cutoff, needed to counter the well-known IR divergence of the 1+1 Klein-Gordon theory of a free massless scalar. Since the cMERA construction reproduces the infrared behaviour of the target state, it also displays such a divergence. Indeed, the integral that defines $f_\varepsilon(x-y)$ can be seen to diverge for $\varepsilon=0$.
		We thus regulate this divergence by introducing an additional IR length scale $1/(\varepsilon\Lambda)$ and removing the degrees of freedom at length scales larger than $1/(\varepsilon\Lambda)$ from the integrals that translate from momentum space back into position space.
		\begin{figure}
			\centering
			\includegraphics[width=0.8\linewidth]{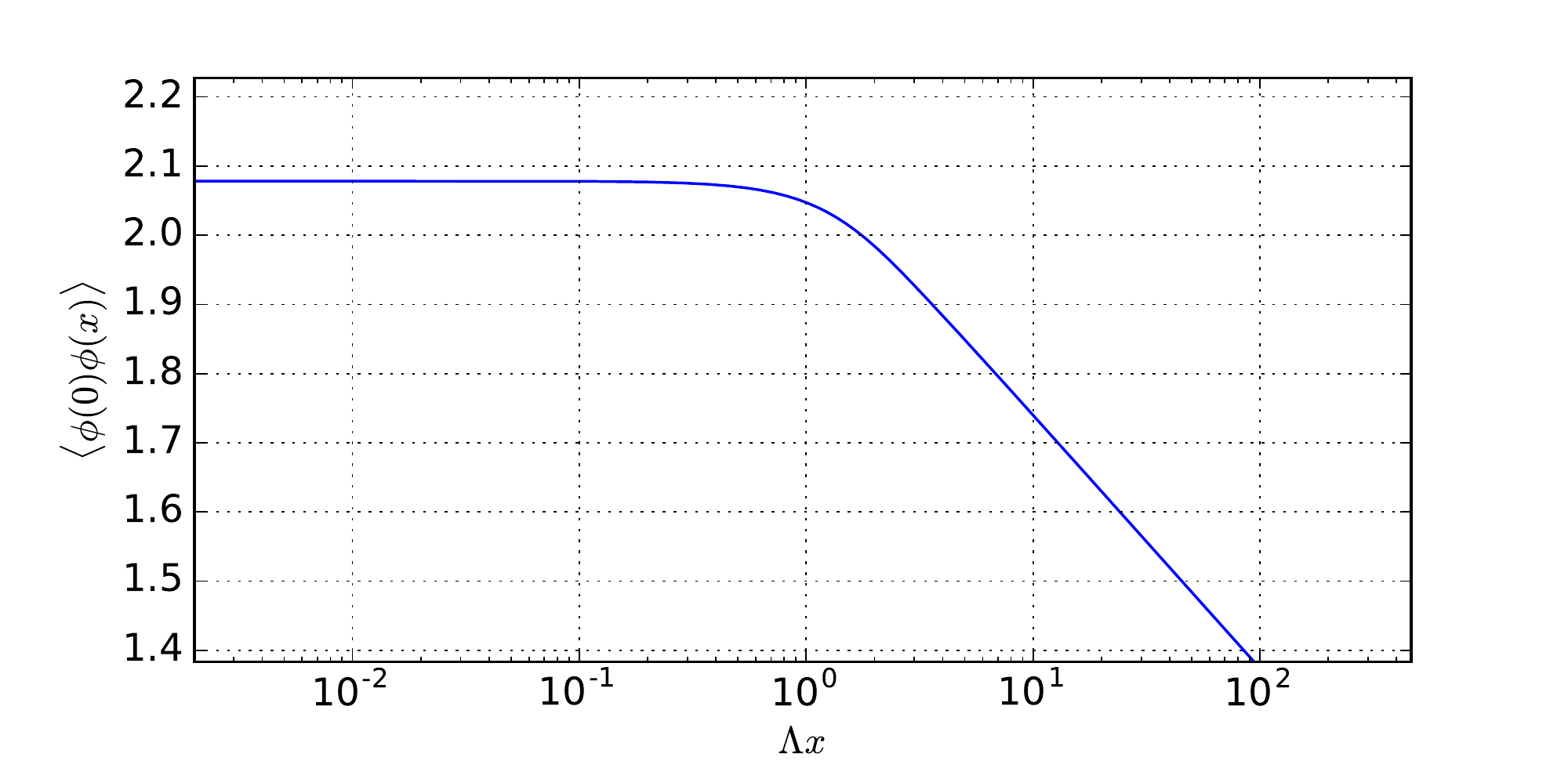}
			\caption{$\la\phi(0)\phi(x)\ra$ correlator computed for a cMERA defined by \ref{micmera}. Notice the existence of two clearly different regimes delimited by $\Lambda x\sim 1$. These results were obtained with an IR regulator with $\varepsilon=10^{-6}$.}
			\label{corr1dbosonsff}
		\end{figure}
		
		\begin{figure}
			\centering
			\includegraphics[width=0.8\linewidth]{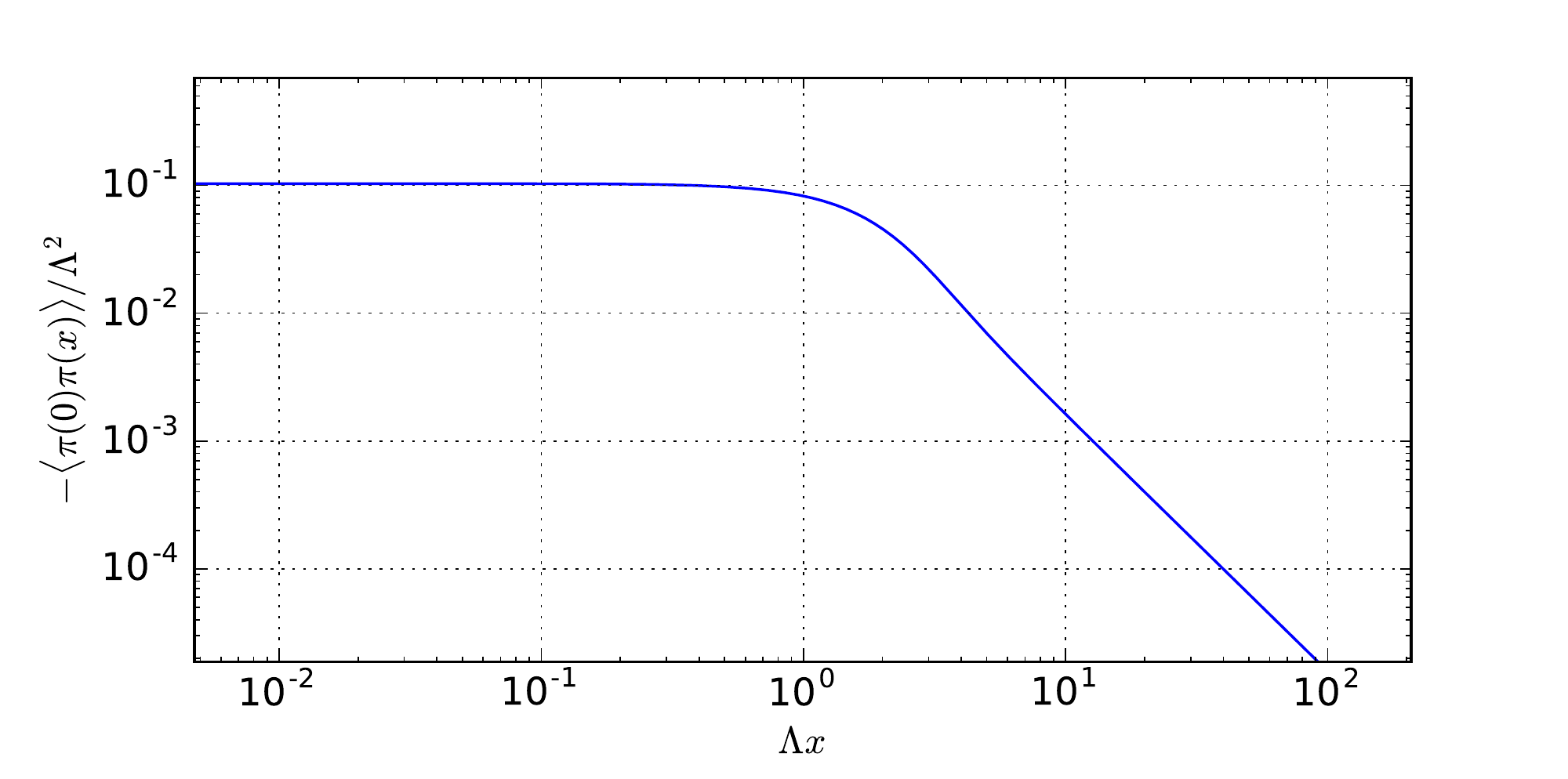}
			\caption{$\la\pi(0)\pi(x)\ra$ correlator computed for a cMERA defined by \ref{micmera}. Notice the existence of two clearly different regimes delimited by $\Lambda x\sim 1$. These results were obtained with an IR regulator with $\varepsilon=10^{-6}$.}
			\label{corr1dbosonspp}
		\end{figure}
Figures \ref{corr1dbosonsff}-\ref{corr1dbosonspp} show the cMERA two-point correlators. Two very different regimes can be appreciated. First, for short distances $x \ll 1/\Lambda$, the correlators are practically constant. In contrast, for distances larger than $1/\Lambda$ (but smaller than the IR cutoff scale given by $1/( \varepsilon\Lambda)$) the correlators recover the scaling expected in the CFT:
	\begin{equation}
	\la\phi(x)\phi(y)\ra=-\dfrac{\log|x-y|}{2\pi}\qquad	\la\pi(x)\pi(y)\ra\propto \dfrac{-1}{(x-y)^2}
	\end{equation}
	From the numerically obtained cMERA correlations in the regime of distances $x$ given by $1/\Lambda \ll x \ll 1/(\varepsilon \Lambda)$,  $\varepsilon=10^{-6}$, we can estimate the following values for the coefficient of the logarithmic decay and the exponent of the power law decay:
	\begin{align}
	\la\phi(x)\phi(y)\ra\sim-p_1\log|x-y|\qquad p_1\approx 0.15904\\
	\la\pi(x)\pi(y)\ra\sim \dfrac{-1}{(x-y)^{p_2}}~~~~~~\qquad p_2\approx 2.0078~
	\end{align}
	which indeed are very close to their values for the target CFT theory, namely $1/(2\pi) = 0.15915$ and $2\Delta_\pi=2$, where $\Delta_\pi=1$ is the scaling dimension of $\pi$. The value of $p_2$ can in fact be obtained from the momentum space representation \eqref{momspa}, prior to the numerics, via asymptotic analysis. This is explained in Appendix \ref{asymfou}. Indeed, the fact that $\alpha(k)$ has a discontinuity in its first derivative at $k=0$ imposes for the two-point function in position space an asymptotic power-law decay of the form $\la\pi(x)\pi(y)\ra\sim|x-y|^{-2}$. This discontinuity in $\alpha(k)$ was already there in the target theory, of which cMERA preserves the low momentum characteristics.
	By the same kind of arguments we can also compute the leading order asymptotic term of the difference between CFT and cMERA correlators. Since
	\begin{equation}
	\alpha_\text{cMERA}(k)-\alpha_\text{CFT}(k) = -\dfrac{k^3}{2\sigma\Lambda^2}+\ldots
	\end{equation}
	we have, at long distances
	\begin{equation}
	|\la\pi(0)\pi(x)\ra_{\text{cMERA}} -\la\pi(0)\pi(x)\ra_{\text{CFT}}| = \dfrac{3}{2\pi\sigma \Lambda^2 x^4}+\ldots
	\end{equation}
	
	\subsubsection*{Entanglement entropy}
	In the CFT, the entanglement entropy of a finite interval is divergent. We can still compute its characteristic scaling in Eq. \ref{ctercioslog} by discretizing the CFT into a lattice, where the lattice spacing $a$ becomes the UV cutoff in Eq. \ref{ctercioslog}. Importantly, progressive fine-graining of the lattice brings in new degrees of freedom that contribute to the entropy, which therefore diverges as the lattice spacing is removed in the limit $a\rightarrow 0$.
	
	Our procedure for numerically computing the entanglement entropy in the cMERA is also based on a lattice discretization. However, instead of discretizing and solving the full theory on the lattice, we will simply sample the continuum two-point correlation functions on a lattice with lattice spacing $a$.
\begin{figure}
	\centering
	\includegraphics[width=0.9\linewidth]{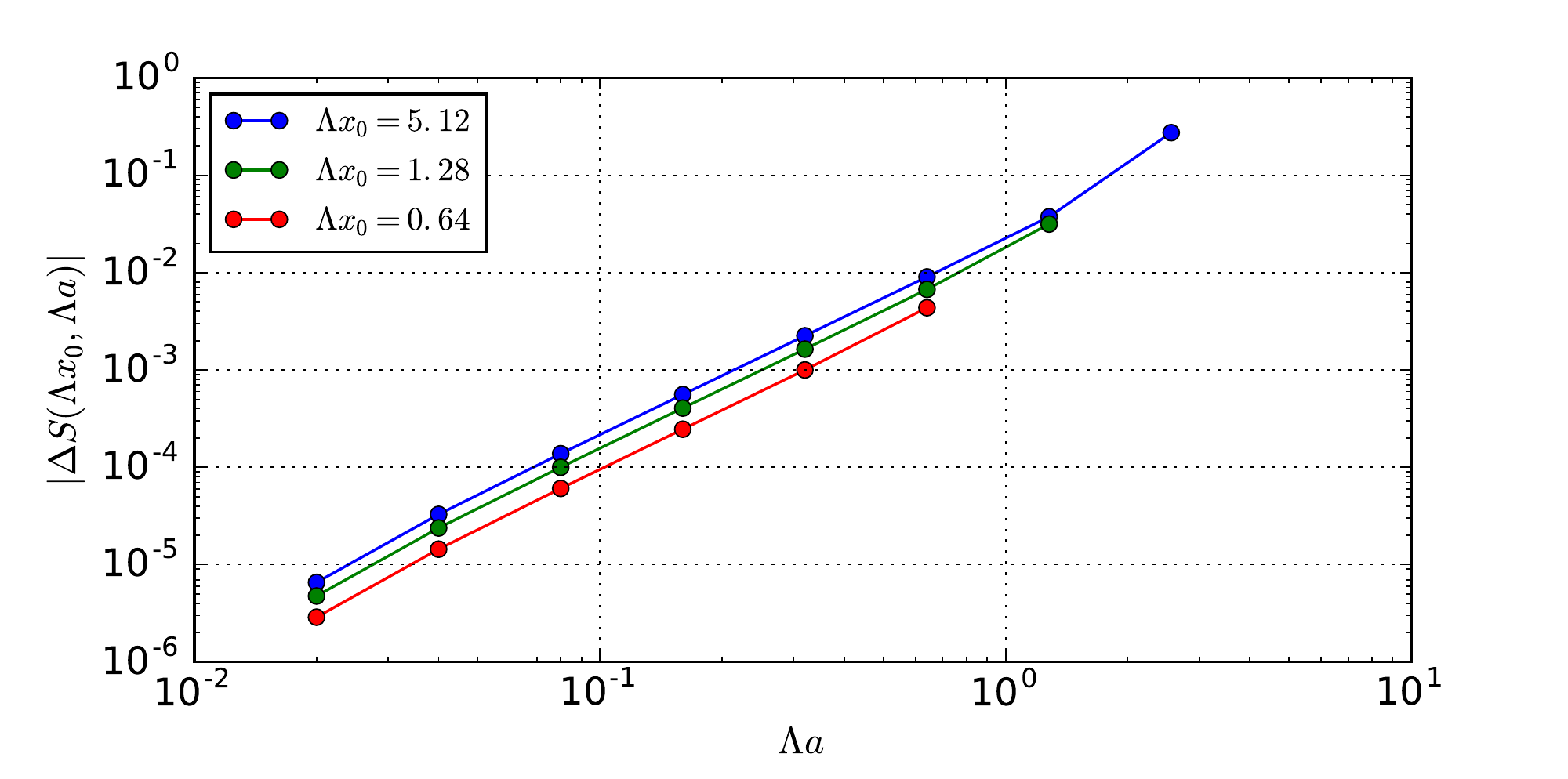}
	\caption{Plot of the difference $|S(\Lambda x_0,\Lambda a)-S(\Lambda x_0,\Lambda a = 0.01)|$ that shows the convergence of this particular value of entropy upon iterative fine-graining of the sampling parameter used as a tool to compute it. The plotted difference goes to zero approximately quadratically with $\Lambda a$.}
	\label{fig:ent1dbosonsconvergence}
\end{figure}
	
This procedure yields discrete matrix versions of the correlation functions $C_{\phi\phi}=\la\phi(x)\phi(y)\ra$ and $C_{\pi\pi}=\la\pi(x)\pi(y)\ra$, from which one can easily extract the entanglement entropy following Appendix \ref{apa}. We apply the following discretization conventions
	\begin{align}
	C_{\phi\phi}(x,y)&=\dfrac{1}{2\Lambda}\delta(x-y)+f_\varepsilon(x,y)&\longrightarrow\; (C_{\phi\phi})_{ij}&=\dfrac{1}{2\Lambda a}\delta_{ij}+f_\varepsilon(ia,ja)\qquad i,j\in\Z\,,\\
	C_{\pi\pi}(x,y)&=\dfrac{\Lambda}{2}\delta(x-y)+g_\varepsilon(x,y)&\longrightarrow\; (C_{\pi\pi})_{ij}&=\dfrac{\Lambda a}{2}\delta_{ij}+a^2g_\varepsilon(ia,ja)\qquad i,j\in\Z\,.
	\end{align}
	In this manner we numerically compute the entanglement entropy profile $S(x)$. We must note that this method is based on an approximation. The actual theory is defined in the continuum, and we are applying a sampling operation that produces discretized versions of the continuous operator kernels $C(x,y)$, in the hope that their spectra will be captured well enough by those of their discretizations in the limit of small $a$. Note that, for example, the spectra of the discretized operators will in general not be fully compatible with the constraints on the spectra of discrete correlation matrices. This forces us to discard a fraction of the eigenvalues of the constructed operators, relying on the assumption that their deviation from allowed values approaches zero as $a$ does (an assumption that we corroborate numerically).  Crucially, progressive fine-graining of the lattice discretization (that is, reducing the lattice spacing $a$) reveals convergence of the entanglement entropy $S(x)$ to a finite value, rather than the divergence seen in the entanglement entropy of the CFT. Specifically, Figure \ref{fig:ent1dbosonsconvergence} shows that for $a \ll 1/\Lambda$, the entanglement entropy of an interval converges to its finite value for $a=0$ quadratically in $a$, $S_a(x)= S(x) + O(a^2)$. We will take this value as our approximation to the entanglement entropy.

\begin{figure}
\centering
\includegraphics[width=0.9\linewidth]{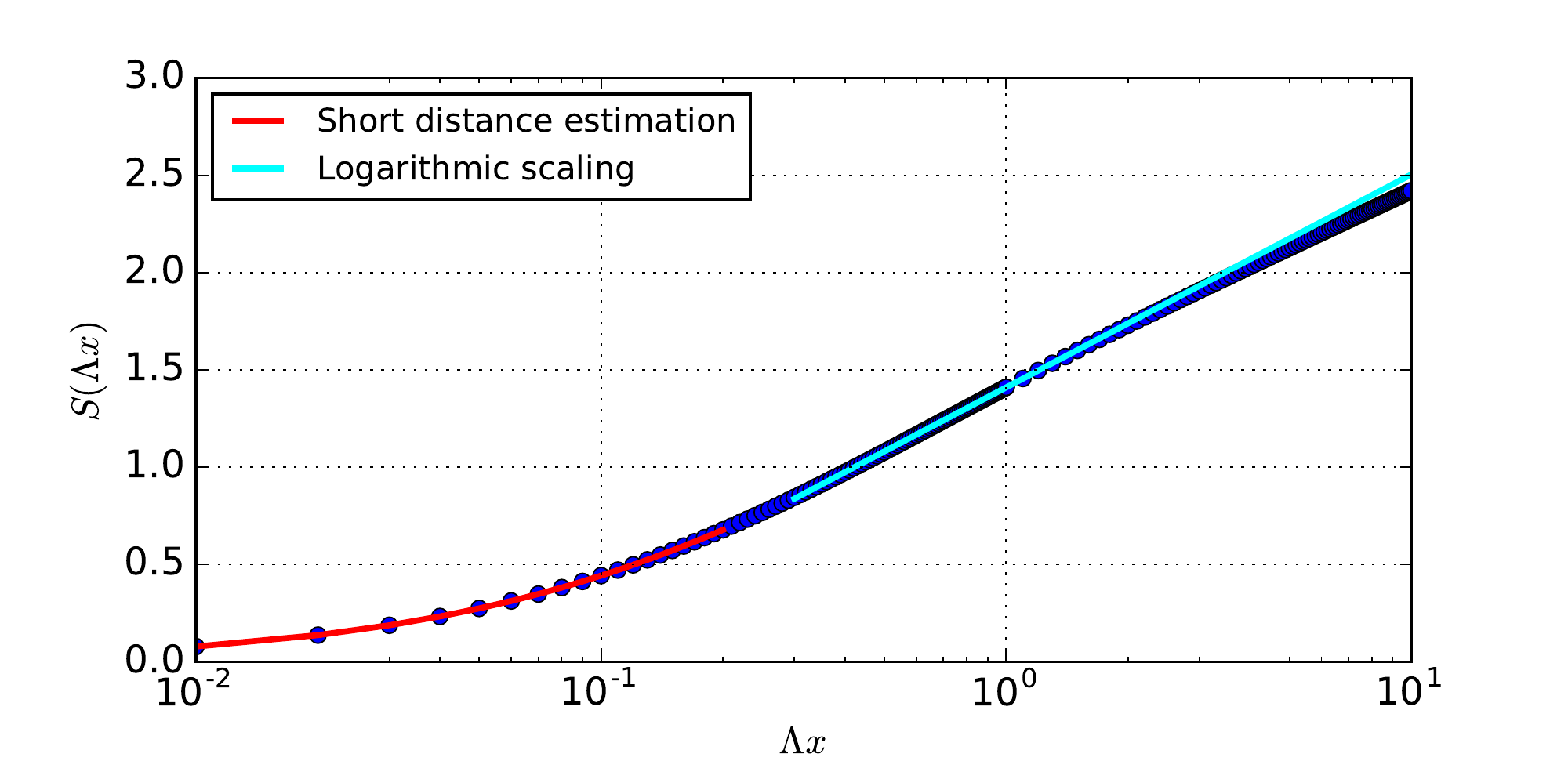}
\caption{Numerical computation of the entanglement entropy profile ($\Lambda a = 0.01,\;\varepsilon=10^{-6}$) for a 1+1 dimensional Klein-Gordon theory. The short range theoretical estimation and the long distance logarithmic scaling have been superimposed.}
\label{ent1dbosons}
\end{figure}

The converged entropy profile can be seen in Figure \ref{ent1dbosons}. We observe two clearly different regimes: one for interval sizes $x$ significantly smaller than the UV cutoff $1/\Lambda$, and one for sizes $x$ comparable to and larger than $1/\Lambda$ (but smaller than the IR cutoff $1/(\varepsilon\Lambda)$, which is not shown). In this second region, the scaling of entanglement entropy reproduces the CFT logarithmic growth of \eqref{ctercioslog}. The numerical fit of the central charge $c$ in the region around $x\sim 1/\Lambda$ gives a value of the central charge
	\begin{equation}
	c\approx 0.987
	\end{equation}
	which is very close to the exact value $c=1$. At larger distances the growth of the entropy is slightly smaller than the one dictated by \eqref{ctercioslog} with $c=1$, which we believe to be an effect of the IR regulator. Figure \ref{ent1dbosons} also shows a theoretical estimate, derived in Appendix \ref{apb}, for the scaling of the cMERA entanglement entropy for a small interval size $x$, $x\ll 1/\Lambda$.
	
	\subsubsection*{Built-in UV cutoff}
	
We have thus seen that the scaling of correlations and of entanglement entropy in the cMERA state $\ket{\Psi^{\Lambda}}$ mimic those of the target CFT ground state at large distances $x\gg 1/\Lambda$. However, at short distances $x \ll 1/\Lambda$, the correlators tend to a constant and the entanglement entropy (which is finite for any finite interval) vanishes. Similar results will be obtained below for $d=2$ spatial dimensions and for free fermion CFTs. This is the sense in which the cMERA state $\ket{\Psi^\Lambda}$ has a built-in UV cutoff at distance $x\approx 1/\Lambda$.

\subsection{Free massless boson in 2+1 dimensions}

In more than one spatial dimension, the free massless boson field theory does not need to be IR regularized. However, the cost of computing the correlations and, especially, the entanglement entropy becomes much larger. Fortunately, this does not prevent us from  being able to numerically characterize the scaling at short distances $x \ll 1/\Lambda$, confirm once more the presence of the UV cutoff, and justify both analytically and numerically the transition to the asymptotic CFT-like behaviour.

\subsubsection*{Two-point correlation functions}
	The cMERA two-point correlation functions can again be computed from the momentum space correlation function, which is written in terms of the function $\alpha(k)$:
		\begin{align}
	\la\phi(\vec k)\phi(\vec q)\ra=\dfrac{1}{2\alpha(k)}\delta( \vec k+\vec q),~~~~~~~~~~~~\la\pi(\vec k)\pi(\vec q)\ra=\dfrac{\alpha(k)}{2}\delta(\vec  k+\vec q),
	\label{momsparep}
	\end{align}
	just by Fourier transform:
	\begin{align}
	\la\phi(\vec x)\phi(\vec y)\ra&=\dfrac{1}{2\Lambda}\delta(\vec x-\vec y)+\int_{\R^2}{d\vec k ~\dfrac{e^{i\vec k \cdot(\vec x-\vec y)}}{(2\pi)^2}\left( \dfrac{1}{2\alpha(k)}-\dfrac{1}{2\Lambda}\right)}\nonumber\\
	&=\dfrac{1}{2\Lambda}\delta(\vec x-\vec y)+\int_{0}^{\infty}{\dfrac{k\:dk}{2\pi} \left( \dfrac{1}{2\alpha(k)}-\dfrac{1}{2\Lambda}\right)\int_{0}^{2\pi}{d\varphi \dfrac{e^{ik |\vec x-\vec y|\cos\varphi}}{2\pi}}}\nonumber\\
	&=\dfrac{1}{2\Lambda}\delta(\vec x-\vec y)+\int_{0}^{\infty}{\dfrac{k\:dk}{2\pi} \left( \dfrac{1}{2\alpha(k)}-\dfrac{1}{2\Lambda}\right)J_0(k |\vec x-\vec y|)}\nonumber\\
	&=\dfrac{1}{2\Lambda}\delta(\vec x-\vec y)+f(|\vec x-\vec y|)\label{f2b}
	\end{align}
	where $J_0$ is the zeroth Bessel function of the first kind. Similarly,
	\begin{align}
	&\la\phi(\vec x)\pi(\vec y)\ra=\dfrac{i\delta(\vec x-\vec y)}{2}=\overline{\la\pi(\vec x)\phi(\vec y)\ra},\\
	&\la\pi(\vec x)\pi(\vec y)\ra=\dfrac{\Lambda}{2}\delta(\vec x-\vec y)+\int_{0}^{\infty}{\dfrac{k\:dk}{2\pi} \left( \dfrac{\alpha(k)-\Lambda}{2}\right) J_0(k |\vec x-\vec y|)}\\
	&\phantom{\la\pi(\vec x)\pi(\vec y)\ra}=\dfrac{\Lambda}{2}\delta(\vec x-\vec y)+g(|\vec x-\vec y|).\label{g2b}
	\end{align}
These two-point correlation functions consist of on-site deltas plus smooth terms\footnote{Notice that we repeat the notation $f,g$ for these terms, which should not be mistaken for their 1+1-dimensional analogues. We will also reuse this notation in the fermionic case. Note as well that we won't need IR regularization outside the 1+1 dimensional bosonic theory.} that only depend on the distance $|\vec x-\vec y|$. Their asymptotic behaviour for long distances can again be inferred from their momentum space representation \eqref{momsparep}. $\la\phi(\vec x)\phi(\vec y)\ra$ decays as $|\vec x-\vec y|^{-1}$ at long distances due to the $|\vec k|^{-1}$ singularity at the origin of the function $[2\alpha(\vec k)]^{-1}$ (see Appendix \ref{asymfou}). On the other hand, $\la\pi(\vec x)\pi(\vec y)\ra$ decays as $|\vec x-\vec y|^{-3}$ since the function $\alpha(\vec k)/2$ goes as $|\vec k|$ near $\vec k=0$. Note again how the fact that the cMERA construction preserves the low momentum character of the target CFT  manifestly leads to the preservation of the long distance behaviour of the two-point functions of the theory. The leading order asymptotic cMERA corrections to these can be seen to go like 
	\begin{align}
	|\la\phi(\vec x)\phi(\vec y)\ra_{\text{cMERA}}-\la\phi(\vec x)\phi(\vec y)\ra_{\text{CFT}}|\sim|\vec x-\vec y|^{-3}\\
	|\la\pi(\vec x)\pi(\vec y)\ra_{\text{cMERA}}-\la\pi(\vec x)\pi(\vec y)\ra_{\text{CFT}}|\sim |\vec x-\vec y|^{-5}
	\end{align}
	 i.e., decaying in both cases two orders faster than the leading term.

\begin{figure}
		\centering
		\includegraphics[width=0.8\linewidth]{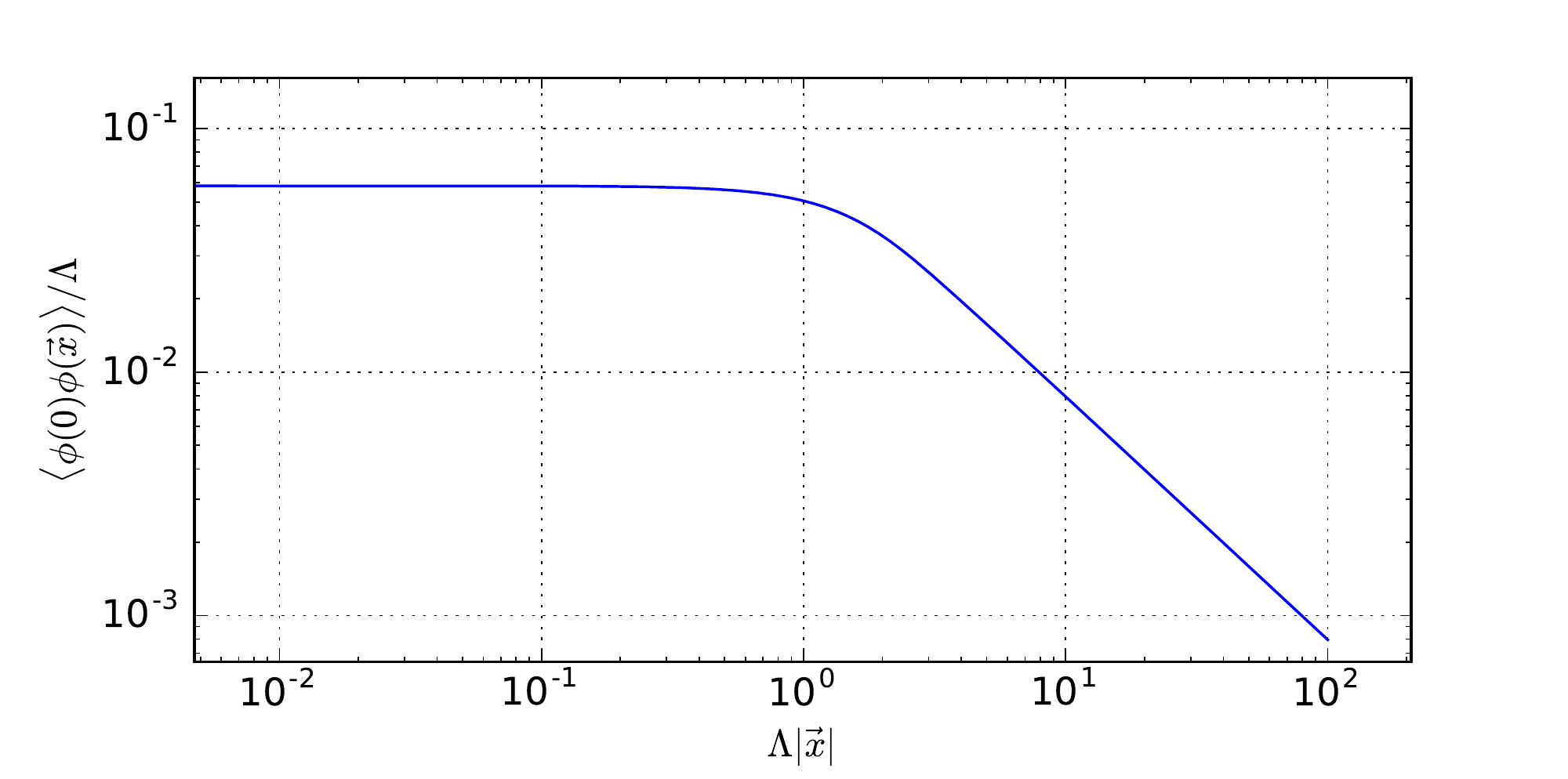}
		\caption{$\la\phi(0)\phi(\vec{x})\ra$ correlator computed for the 2+1-dimensional bosonic cMERA. }
		\label{corr2dbosonsff}
	\end{figure}

\begin{figure}
			\centering
			\includegraphics[width=0.8\linewidth]{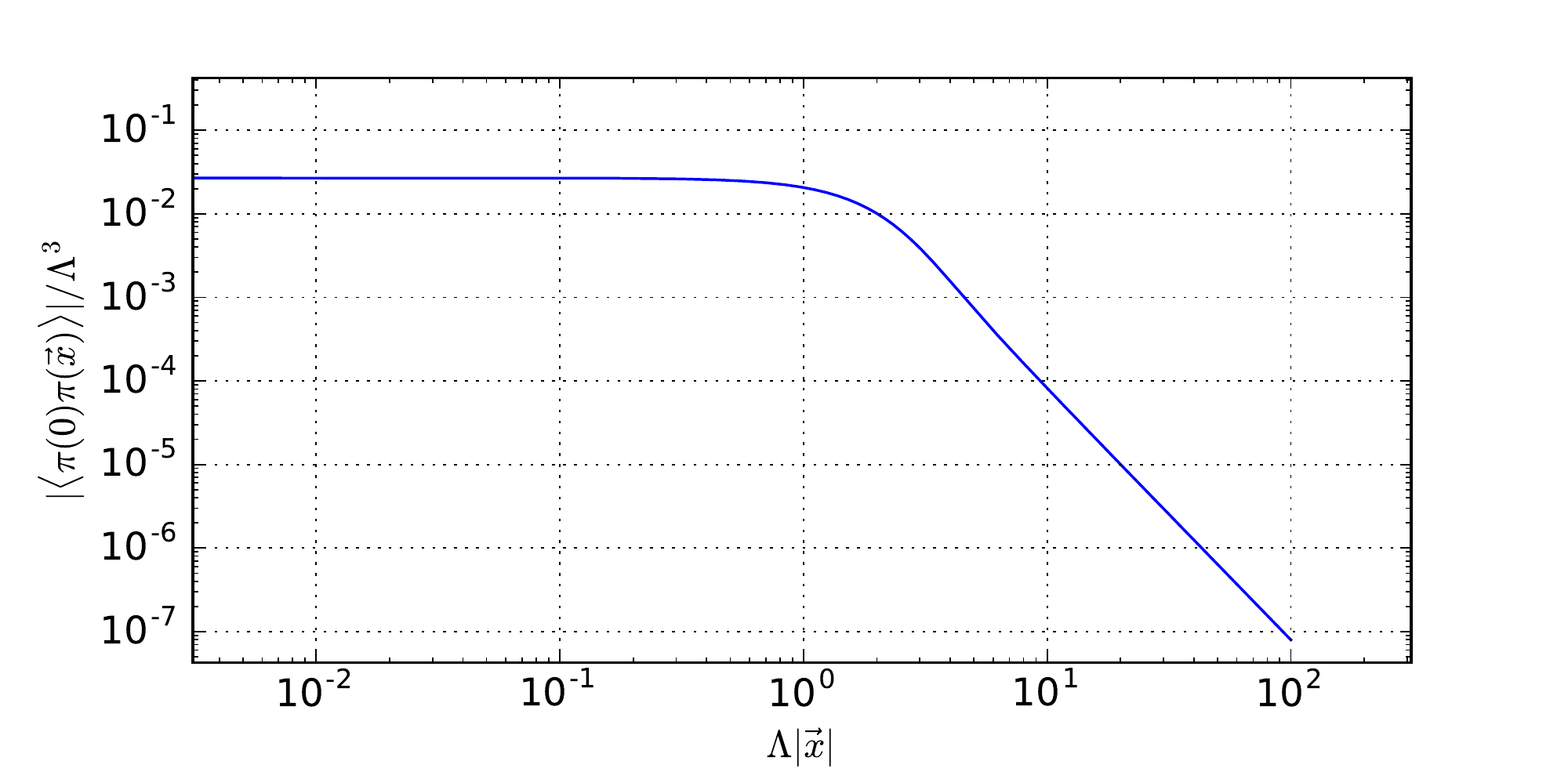}
			\caption{$\la\pi(0)\pi(\vec{x})\ra$ correlator computed for the 2+1-dimensional bosonic cMERA. }
			\label{corr2dbosonspp}
	\end{figure}

We plot the numerically obtained functions $f$ and $g$ in Figures \ref{corr2dbosonsff}-\ref{corr2dbosonspp}. As in the one dimensional case, we observe a short distance regime $|\vec x-\vec y|\ll 1/\Lambda$, where the correlators are practically constant, and a long distance regime $|\vec x-\vec y|\gg 1/\Lambda$, where the shape of the correlators reproduces the CFT power law decay with the right exponents:
	\begin{align}
	f(x)\sim \frac{1}{x^{2p_1}},\qquad p_1=0.4998 ~~~(\approx 0.5 =\Delta_{\phi}),\qquad\Lambda x\gg 1,\\
	g(x)\sim \frac{1}{x^{2p_2}},\qquad p_2=1.502 ~~~~~(\approx 1.5 =\Delta_{\pi}), \qquad\Lambda x\gg 1.
	\end{align}
	\subsubsection*{Entanglement entropy}
	In $2+1$ dimensions we compute the entanglement entropy of discs of increasing radius $x$. The technical details of this computation are presented in Appendix \ref{apc}. In short, we work in polar coordinates and consider modes indexed by the radial coordinate $r$ and with a definite angular momentum given by an integer $l\in\Z$. Different angular momentum modes are uncorrelated, so they contribute independently to the entanglement entropy. Only the modes with smallest angular momentum are found to contribute at short distances, with the corrections due to larger angular momenta becoming more relevant at longer distances.

\begin{figure}
			\centering
			\includegraphics[width=0.9\linewidth]{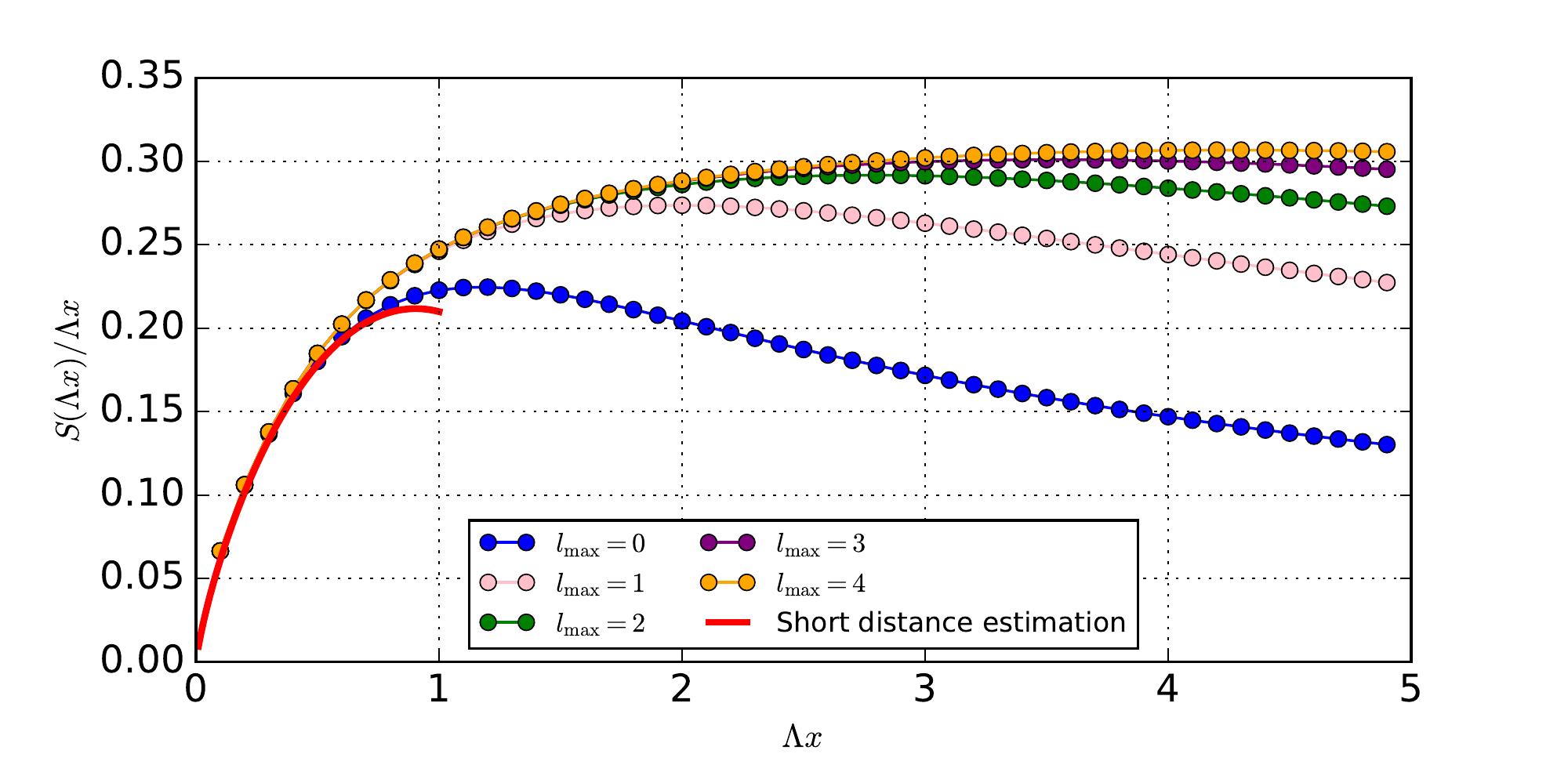}
			\caption{Numerical computation of the entanglement entropy profile ($\Lambda a = 0.01$) for the 2+1 dimensional bosonic cMERA.}
			\label{fig:ent2dbosons}
\end{figure}
		
\begin{figure}
	\centering
	\includegraphics[width=0.9\linewidth]{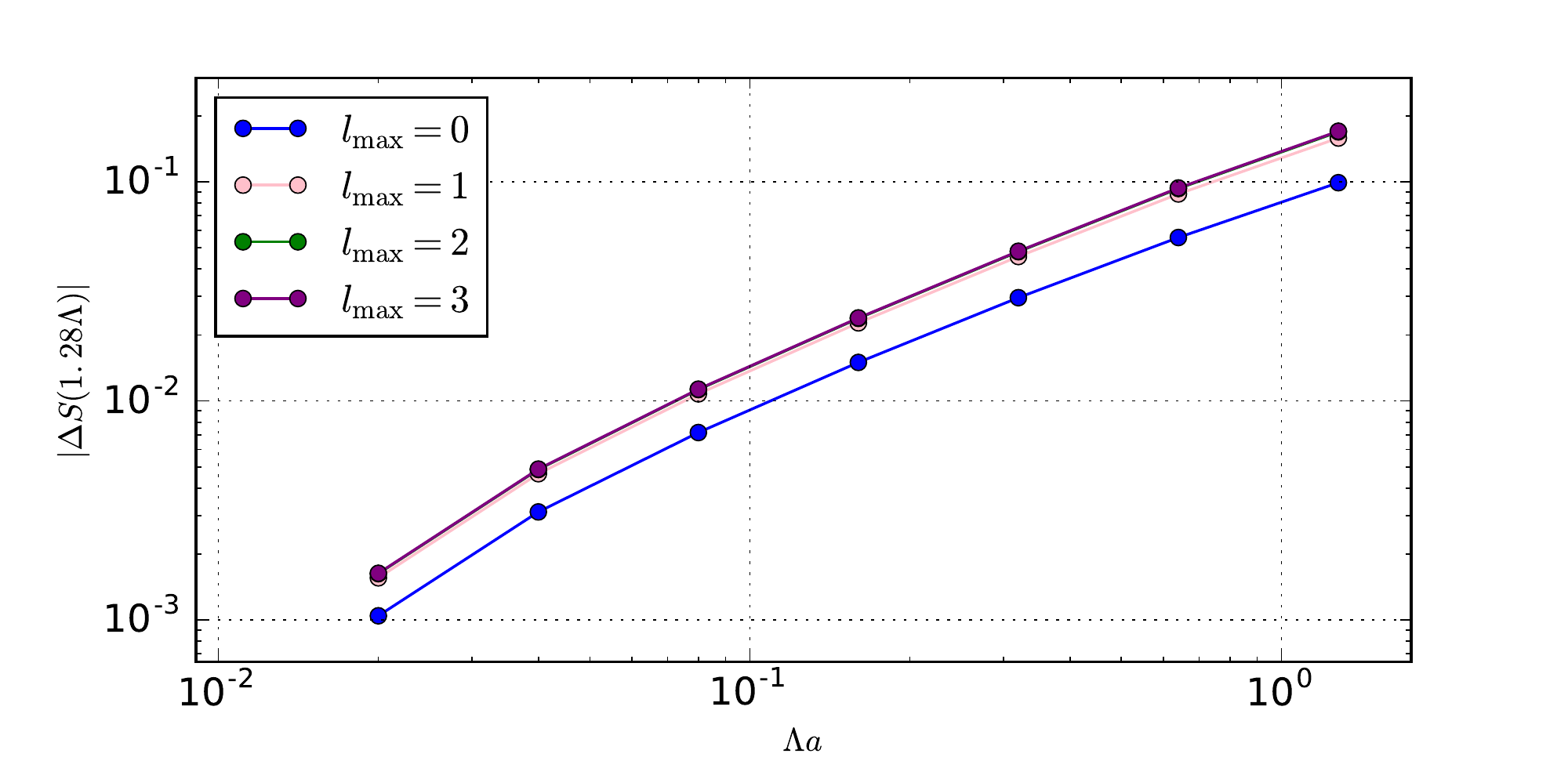}
	\caption{Plot of the difference $|S(1.28\Lambda,\Lambda a)-S(1.28\Lambda,\Lambda a = 0.01)|$, for different values of $l_\text{max}$ that shows the convergence of this particular value of entropy upon iterative fine-graining of the sampling used as a tool to compute it. Note how truncating at higher values of $l$ does not affect the rate of convergence in $a$.}
	\label{fig:ent2dbosonsconvergence}
\end{figure}

	We approximately compute the entanglement entropy by the same procedure as in the 1+1-dimensional case: we sample the correlators with some lattice spacing $a$ and build discrete versions of the continuum operators $C_{\phi\phi}, C_{\pi\pi}$, from which we numerically extract an approximation to the entanglement entropy $S(x)$. In addition, Appendix \ref{apb} derives a theoretical estimate of $S(x)$ for $x \ll 1/\Lambda$, restricted to the contribution from the zero angular momentum mode.

	Two remarks are in order. The first refers to the convergence of the profile $S(x)$ above with respect to contributions coming from different angular momenta. Figure \ref{fig:ent2dbosons}, shows the partial entropy $S_{l_{\max}}(x)$ of a disc of radius $x$ obtained by adding the contributions from all angular momentum $l$ such that $|l| \leq l_{\max}$. We see that, indeed, $S_{l_{\max}}(x)$ is essentially independent of $l_{\max}$ for $x \leq 1/\Lambda$. For $x \approx 2/\Lambda$ convergence is roughly obtained for $l_{\max} \geq 1$, for $x\approx 3/\Lambda$ convergence is roughly obtained for $l_{\max} \geq 2$, etc. These results are sufficient to see the outset of the area law at $x\approx 1/\Lambda$, as expected in the CFT, see Eq. \ref{arealaw}.
	
	Our second remark refers to the convergence of these results with respect to the sampling parameter $a$ used to discretize the correlation functions. Figure \ref{fig:ent2dbosonsconvergence} shows that, once more, $S(x)$ tends to a finite profile when we reduce $a$. Notice also the agreement between the numerical values and the zero angular momentum estimate at short distances $x \ll 1/\Lambda$.

	\section{Free fermionic QFTs}
	\label{sec4}
	In this section we investigate the short-distance entanglement structure of the cMERA for the ground state of free fermion CFTs in $1+1$ and $2+1$ spacetime dimensions.
	 Throughout this section, $\vec{\psi}(\vec{x})$ denotes a 2-component Dirac spinor, with components $\psi_1(x)$ and $\psi_2(x)$ obeying anticommutation relations $\left\{\psi^{\phantom{\dagger}}_i(\vec{x}),\psi^\dagger_j(\vec{y}) \right\} = \delta_{i,j}\delta(\vec{x}-\vec{y})$ and $\left\{\psi_i^{\phantom{\dagger}}(\vec{x}),\psi_j^\dagger(\vec{y}) \right\} = 0$. Similarly, $\vec{\psi}(\vec{k})$ denotes the Fourier component of the Dirac spinor, with  $\left\{\psi^{\phantom{\dagger}}_i(\vec{k}),\psi^\dagger_j(\vec{q}) \right\} = \delta_{i,j}\delta(\vec{k}-\vec{q})$ and $\left\{\psi^{\phantom{\dagger}}_i(\vec{k}),\psi^\dagger _j(\vec{q})\right\} = 0$, where
	 \begin{equation}
	 \psi_i(\vec{k}) \equiv \frac{1}{(2\pi)^{d/2}} \int d^dx~ e^{-i\vec{k}\cdot \vec{x}}  \psi_i(\vec{x}),~~~~i=1,2.
	 \end{equation}

	\subsection{Fermionic Gaussian cMERA framework}
	The Gaussian states $\ket{\Phi}$ under consideration are annihilated by annihilation operators $\tilde{\psi}_1(\vec{k})$ and $\tilde{\psi}_2^{\dagger}(\vec{k})$,
	\begin{equation}
	\tilde \psi_1(\vec k)\ket{\Phi}=0,~~~~~\tilde{\psi}_2^\dagger(\vec k)\ket{\Phi}=0,~~~~\forall \vec{k} \in \mathbb{R}^d,
	\end{equation}
where $\tilde{\psi}_1(\vec{k})$ and $\tilde{\psi}_2(\vec{k})$ are related to the original spinor components $\psi_1(\vec{k})$ and $\psi_2(\vec{k})$ by a $\vec k$-dependent unitary transformation $M(\vec k)$, so that $\tilde{\psi}_i(\vec k)=M_{ij}(\vec k)\psi_j(\vec k)$. For all the states of interest in this section, this transformation can be parameterized by an angular function $\theta(\vec{k})=\theta(k)$ of the momenta that depends on $k\equiv |\vec{k}|$, 
as well as the product $\vec{\gamma}\cdot \hat{k}$, according to 
	\begin{equation}
	M(\vec k)=\cos{\theta(k)}~\mathds{1}+\sin{\theta(k)}~\vec{\gamma}\cdot\hat k=\exp \left(~\theta(k)~\vec{\gamma}\cdot\hat k ~\right). 
	\label{eme}
	\end{equation}
	Here $\hat k \equiv \frac{\vec k}{|\vec k|}$ is a normalized vector and $\vec\gamma$ is the vector of space-like Dirac matrices, whose components $\gamma^i$, to be introduced later on, satisfy $(\gamma^i)^{\dagger} = - \gamma^{i}$.
	
	\subsubsection*{Target state $\ket{\Psi}$}
	Consider a free massless Dirac fermion Hamiltonian in $d=1,2$ spatial dimensions:
	\begin{equation}
	H=\int{d^dx\;\psi^\dagger(\vec x)\gamma^0(-i\vec\gamma\cdot\vec\partial)\psi(\vec x)},
	\end{equation}
	where $\gamma^0$ is the time-like Dirac matrix, which we choose to be
	\begin{equation}
\gamma^0\equiv\left(\begin{array}{cc}
1 & 0\\0 & -1
\end{array} \right).
\end{equation}
In terms of the Fourier space operators,
	\begin{equation}
		H=\int{d^dk\;\psi^\dagger(\vec k)~\gamma^0\left( \vec\gamma\cdot\vec k \right)\psi(\vec k)},
	\end{equation}
the Hamiltonian consists, for each momentum $\vec{k}$, of a quadratic form $\gamma^0(\vec\gamma\cdot\vec k)$ that can be seen to be Hermitian and can thus be diagonalized by a unitary transformation, which turns out to be of the form \eqref{eme}, for 
\begin{equation}
	\theta(k)=\pi/4 ~~~~~~~~~~\left(\mbox{target CFT ground state} ~\ket{\Psi} \right).
\end{equation}	
Indeed,
	\begin{equation}
	\tilde{\psi}(\vec k)=\dfrac{\mathds{1}+\vec{\gamma}\cdot\hat k}{\sqrt 2}\psi(\vec k)\implies H=\int{d^dk\;k\;\tilde\psi^\dagger(\vec k)\gamma^0\tilde\psi(\vec k)}.
	\end{equation}
	
		Thus in that basis we have already diagonalized the Hamiltonian and we can identify the structure of the ground state $\ket\Psi$, which shall be annihilated by the operators $\tilde{\psi}_{1}(\vec{k})$ and  $\tilde{\psi}^\dagger_2(\vec{k})$, that is, the operators that annihilate particles with positive energy and create particles with negative energy. 
		
				\subsubsection*{Product state $\ket{\Omega}$}
		Again following \cite{Haegeman2011} we choose the product state
		\begin{equation}
		\psi_1(\vec x)\ket{\Omega}=0,~~~\psi_2^\dagger(\vec x)\ket{\Omega}=0,~~~ \forall \vec{x},
		\end{equation}
		which Fourier transforms to
		\begin{equation}
		\psi_1(\vec k)\ket{\Omega}=0,~~~\psi_2^\dagger(\vec k)\ket{\Omega}=0,~~~ \forall \vec{k},
		\end{equation}
		and it is thus characterized by
\begin{equation}
	\theta(k)=0 ~~~~~~~~~~\left(\mbox{initial product state} ~\ket{\Omega} \right).
	\end{equation}

		\subsubsection*{cMERA $\ket{\Psi^{\Lambda}}$}
		We consider the cMERA presented in \cite{Haegeman2011} and its higher dimensional analogue, characterized by
		\begin{align}
		L&=i\int {d^dk\;\psi^\dagger_j(\vec k)\left(\vec k\cdot\vec{\nabla}+\dfrac{1}{2}\right)\psi_j(\vec k)},\\
		K&=i\int {d^dk\;g(k)\vec\psi^\dagger(\vec k)\left(\vec{\gamma}\cdot\hat k\right)\vec{\psi}(\vec k)}.
		\end{align}
		We focus our attention on a family of Gaussian cutoff functions:
		\begin{equation}
		g(k)=C_j\left(\dfrac{k}{\Lambda}\right)^{2j+1}e^{-\frac{k^2}{\Lambda^2}},
		\label{ge}
		\end{equation}
		where $j$ is a nonnegative integer and $C_j$ is a constant that depends on $j$. Starting from the product state $\ket{\Omega}$, the result of the evolution in Eq. \ref{inimera} by $L+K$ evolution is the Gaussian state $\ket{\Psi^{\Lambda}}$ characterized by
		\begin{equation}
		\theta(k)=C_j\int_{k/\Lambda}^{\infty}{dz\;z^{2j} e^{-z^2}}=\dfrac{C_j}{2}\;\Gamma\left(j+\dfrac{1}{2},\left(\dfrac{k}{\Lambda}\right)^2 \right),
		\end{equation}
		where $\Gamma(a,x)$ is the upper incomplete gamma function. We choose \begin{equation}
		C_j=\dfrac{\pi}{4}\dfrac{2}{\Gamma\left(j+\dfrac{1}{2}\right)}=\dfrac{2^{n-1}\sqrt{\pi}}{(2n-1)!!},
		\end{equation}
		 so that $\theta(0)=\pi/4$, which is a sensible condition if we want the cMERA constraints to be similar to the target state's at small $k$, and obtain
		\begin{align}
		\theta(k)&=\dfrac{\pi}{4}\Gamma\left(j+\dfrac{1}{2},\left(\dfrac{k}{\Lambda}\right)^2 \right) \Gamma\left(j+\dfrac{1}{2} \right) ^{-1}=\\&=\dfrac{\pi}{4}\left( 1-\text{erf}\left(\dfrac{k}{\Lambda}\right) \right)+\dfrac{\sqrt{\pi}}{4}e^{-k^2}\sum_{n=1}^{j}{\dfrac{2^nk^{2n-1}}{(2n-1)!!}}~~~~\left(\mbox{optimized cMERA state} ~\ket{\Psi^{\Lambda}} \right).
		\label{titak}
		\end{align}
Thus we observe a similar interpolating character as the one found in the bosonic case. In what follows we will focus on the simplest case $j=0$.

	\begin{figure}
		\centering
		\includegraphics[width=0.5\linewidth]{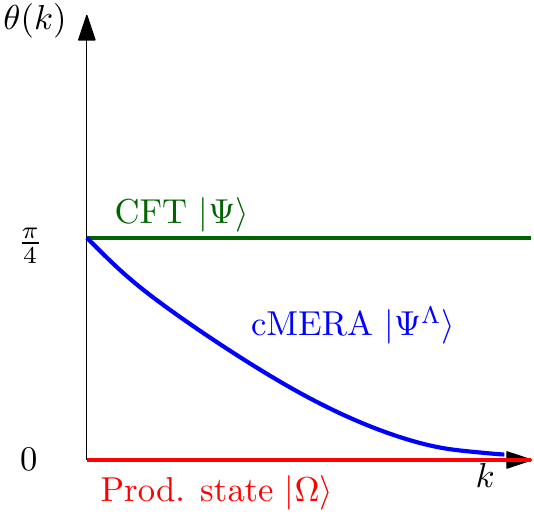}
		\caption{Qualitative plot of the characteristic function $\theta(k)$ for the three states in the fermionic cMERA construction. The behaviour shown for the cMERA is closest to the one obtained by choosing $j=0$.}
		\label{theta}
	\end{figure}
Let us now explore the entanglement structure of this cMERA for 1+1 and 2+1 dimensional theories.

\subsection{Free massless fermion in 1+1 dimensions}
We use the two-dimensional representation of the Dirac $\gamma$ matrices given by
\begin{equation}
\gamma^0=\left(\begin{array}{cc}
1 & 0\\0 & -1
\end{array} \right),\qquad\gamma^1=\left(\begin{array}{cc}
0 & 1\\-1 & 0
\end{array} \right).
\end{equation}
The unitary defining the cMERA is then a SO(2) rotation
\begin{equation}
\left( \begin{array}{c}
\tilde\psi_1(k)\\
\tilde\psi_2(k)
\end{array}\right)=\left( \begin{array}{cc}
	\cos{\theta(k)} & \sign(k)\sin{\theta(k)}\\
	-\sign(k)\sin{\theta(k)} & \cos{\theta(k)}
	\end{array}\right)\left( \begin{array}{c}
	\psi_1(k)\\
	\psi_2(k)
	\end{array}\right).
	\end{equation}
	
\subsubsection*{Two-point correlation functions}

In the CFT ground state, the non-trivial two-point correlation functions are
\begin{align}
\la\psi^\dagger_i(x)\psi_j(y)\ra\propto\begin{cases}
\delta(x-y), & i=j,\\\dfrac{1}{x-y}, & i\neq j.
\label{itmatches}
\end{cases}
\end{align} 
In the cMERA state, the correlation functions in momentum space as a function of the parameter $\theta(k)$ read:
\begin{align}
&\la\psi_1^\dagger(k)\psi^{\phantom{\dagger}}_1(q)\ra=\sin(\theta(k))\sin(\theta(q))\la\tilde\psi_2(k)^\dagger\tilde\psi_2(q)\ra=\sin^2(\theta(k))\delta(k-q)\label{momfer1}\\
&\la\psi_1^\dagger(k)\psi^{\phantom{\dagger}}_2(q)\ra=-\dfrac{1}{2}\sign(k)\sin(2\theta(k))\delta(k-q)=\la\psi_2^\dagger(k)\psi^{\phantom{\dagger}}_1(q)\ra\\
&\la\psi_2^\dagger(k)\psi^{\phantom{\dagger}}_2(q)\ra=\cos^2(\theta(k))\delta(k-q)=\delta(k-q)-\la\psi_1^\dagger(k)\psi^{\phantom{\dagger}}_1(q)\ra
\label{momfer2}
\end{align}
which through a Fourier transform yield their position space counterparts:
\begin{align}
&\la\psi_1^\dagger(x)\psi^{\phantom{\dagger}}_1(y)\ra=\int_{-\infty}^{\infty}{\dfrac{dk}{2\pi}e^{-ik(x-y)}\sin^2(\theta(k))}\label{corrposfer1}\\
&\la\psi_1^\dagger(x)\psi^{\phantom{\dagger}}_2(y)\ra=-\int_{-\infty}^{\infty}{\dfrac{dk}{4\pi}e^{-ik(x-y)}\sign(k)\sin(2\theta(k))}=\overline{\la\psi_2^\dagger(y)\psi^{\phantom{\dagger}}_1(x)\ra}\label{corrposfer2}\\
&\la\psi_2^\dagger(x)\psi^{\phantom{\dagger}}_2(y)\ra=\delta(x-y)-\la\psi_1^\dagger(x)\psi^{\phantom{\dagger}}_1(y)\ra.
\label{corrposfer3}
\end{align}

Before we plot the two-point functions in real space we can obtain information about their long-distance behaviour from their momentum space representation (\ref{momfer1}~--~\ref{momfer2}), as we did for the boson theories (see Appendix \ref{asymfou}). The single-species correlators $\la\psi^\dagger_i(x)\psi_i(y)\ra$, $i=1,2$, will display a leading decay of $(x-y)^{-(2j+2)}$ at long distances. This is due to the first discontinuous derivative of the functions $\sin^2(\theta(k)),\cos^2(\theta(k))$ being the ${(2j+1)\text{-th}}$, which is discontinuous at the origin. This discontinuity was \emph{not} present in the target state, and is the only example in this paper, together with its 2-dimensional counterpart, of a cMERA qualitatively differing from the target state at long distances. The spurious power-law decay of this correlator, not present in the CFT (where the correlator vanished identically everywhere but at the origin) is a consequence of this discontinuity, and can be made faster by choosing a higher $j$, which implies that the resulting $\theta(k)$ coincides to higher and higher order with the exact one in the $k\to 0$ limit \cite{Haegeman2011}.
On the other hand, the dominant term in the asymptotic expansion of the two-species correlators $\la\psi_i^\dagger(x)\psi_j(y)\ra$, $i\neq j$ can be seen to decay like $(x-y)^{-1}$. This reflects the fact that the function $\sign(k)\sin(2\theta(k))$ is discontinuous at the origin, independently of the value of $j$. Looking at \eqref{itmatches}  we see that this indeed matches the behaviour of the target state, and is consistent with the scaling dimension of the fermion fields being $\frac{1}{2}$. What does depend on $j$ is the leading decay order of the difference between the CFT and cMERA correlators:
\begin{equation}
|\la\psi_1^\dagger(x)\psi^{\phantom{\dagger}}_2(y)\ra_{\text{cMERA}}-\la\psi_1^\dagger(x)\psi_2^{\phantom{\dagger}}(y)\ra_{\text{CFT}}|=\dfrac{(4j+2)!\pi}{16\Lambda^{4j+2}\Gamma\left(\frac{3}{2}+j\right)^2 }|x-y|^{-(4j+3)}+\ldots
\end{equation}
The higher the value of $j$, the faster this difference decays, as expected.

\begin{figure}
			\centering
			\includegraphics[width=0.8\linewidth]{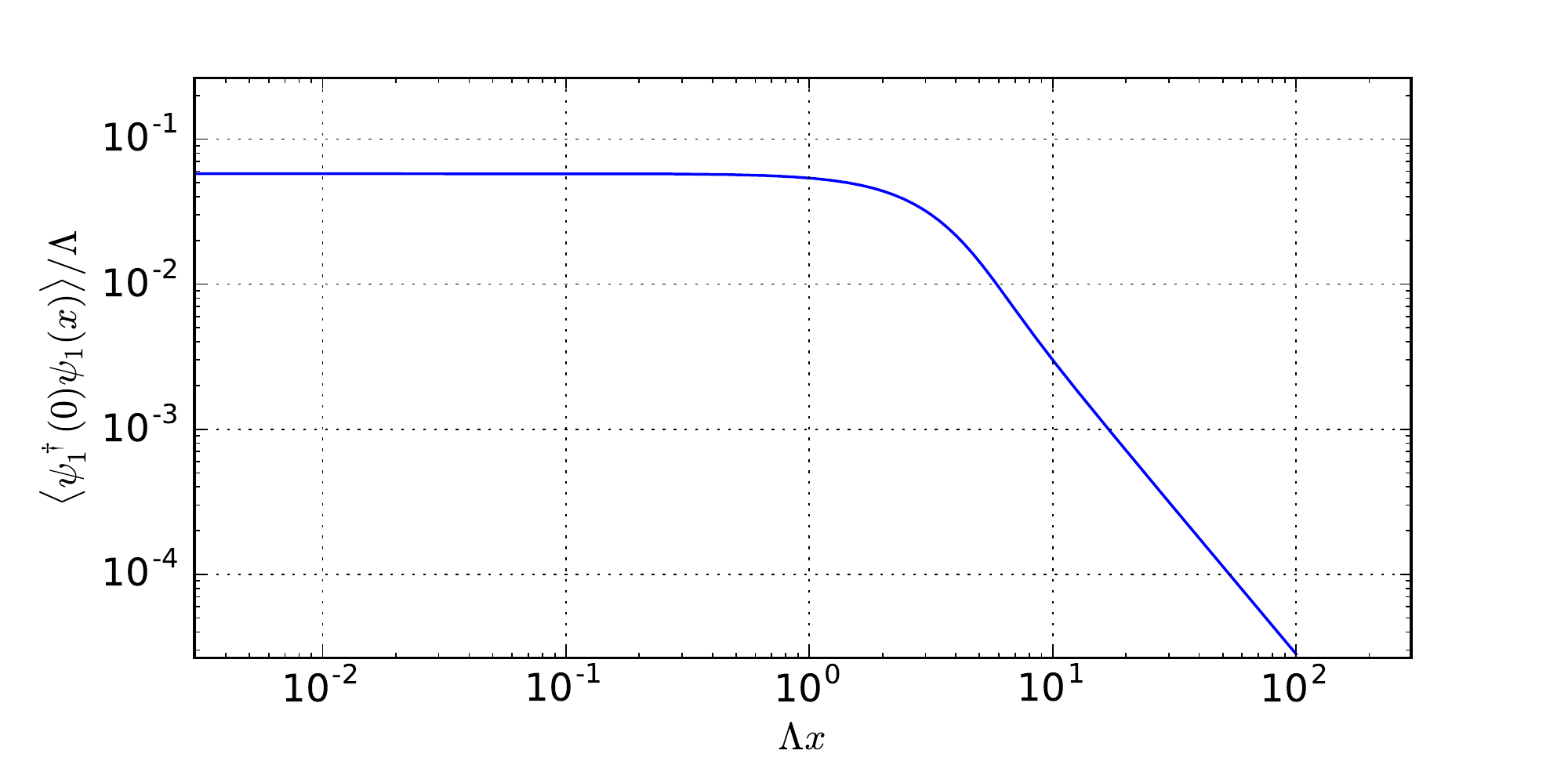}
			\includegraphics[width=0.8\linewidth]{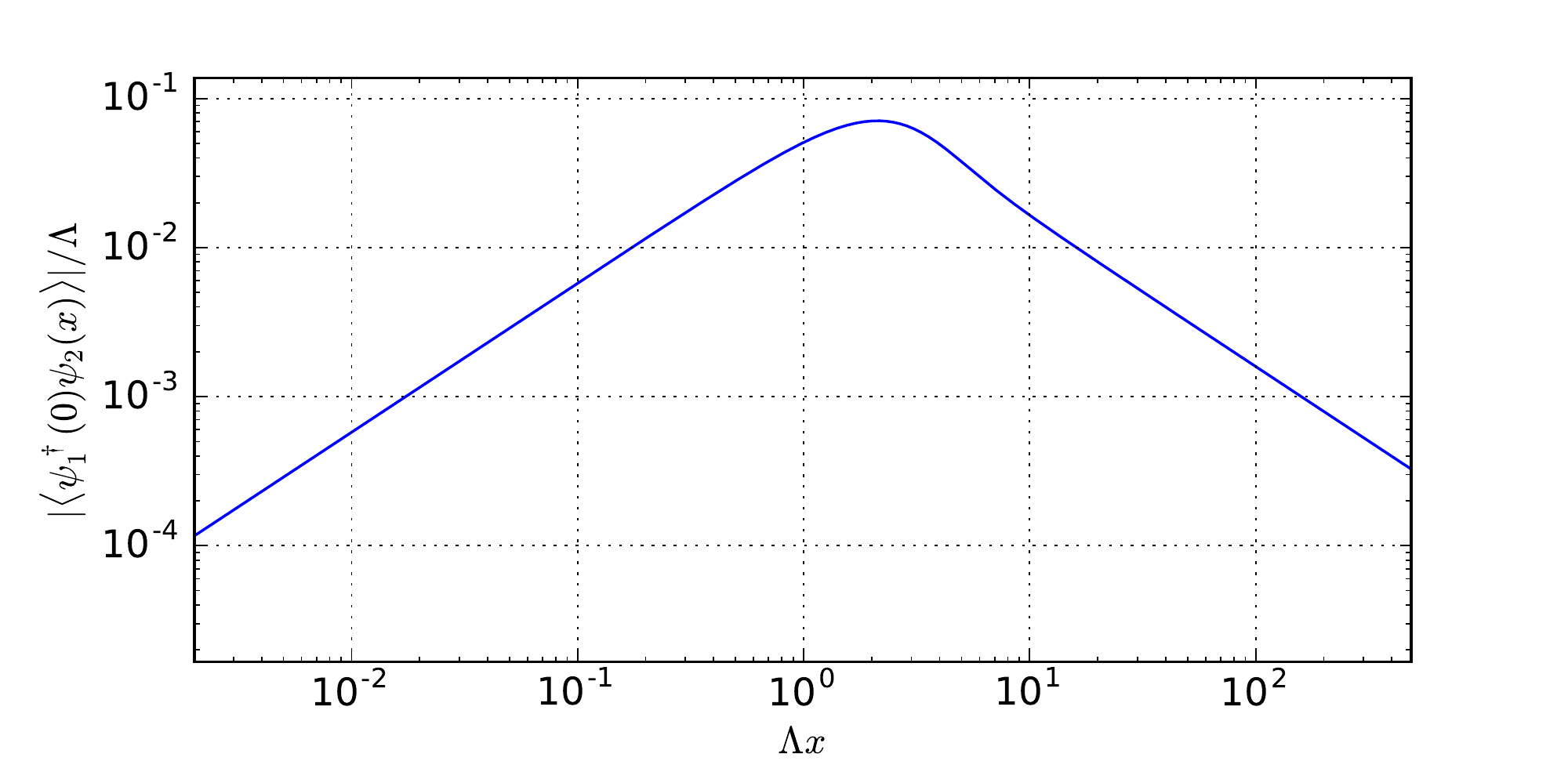}
			\caption{Correlation functions $\la\psi_1^\dagger(0)\psi_1(x)\ra$ (top) $\la\psi_1^\dagger(0)\psi_2(x)\ra$ (bottom, in absolute value) and computed for a cMERA defined by $\theta(k)$ as in \eqref{titak} for $j=0$. Notice the existence of two clearly different regimes delimited by $\Lambda x\sim 1$.}
			\label{corr1dfermions}
		\end{figure}

Fig. \ref{corr1dfermions} displays the numerically computed correlation functions. Only two of them are shown since the others can easily be inferred from those using Eqs. \eqref{corrposfer2} and \eqref{corrposfer3}. At short distances $x \ll 1/\Lambda$, the single-species correlator goes to a constant while the two-species correlator vanishes as $(x-y)^p$, where the exponent is numerically estimated to be $p=0.9992$ (the correlator vanishes linearly as $x-y\rightarrow 0$). 

At large distances $x \gg 1/\Lambda$, both the single-species and two-species correlators exhibit power-law decay, with exponents numerically determined to be $p=-2.002$ and $p=-1.004$ respectively, confirming the previous asymptotic analysis.

		\subsubsection*{Entanglement entropy}
The scaling of the entropy of an interval of length $L$ for the target state is given by \cite{Calabrese2004}:
\begin{equation}
S(x)\sim \dfrac{c}{3}\log\left(\dfrac{x}{a}\right)
\end{equation}
where $c=1/2+1/2=1$, adding the contributions of the two fermionic species in the theory. As in the 1+1 bosonic case, here $a$ is a UV cutoff and the entanglement entropy diverges when $a\to0$. The product state, being devoid of any correlation, displays zero entanglement entropy independently of the region we trace out.

\begin{figure}
			\centering
			\includegraphics[width=0.9\linewidth]{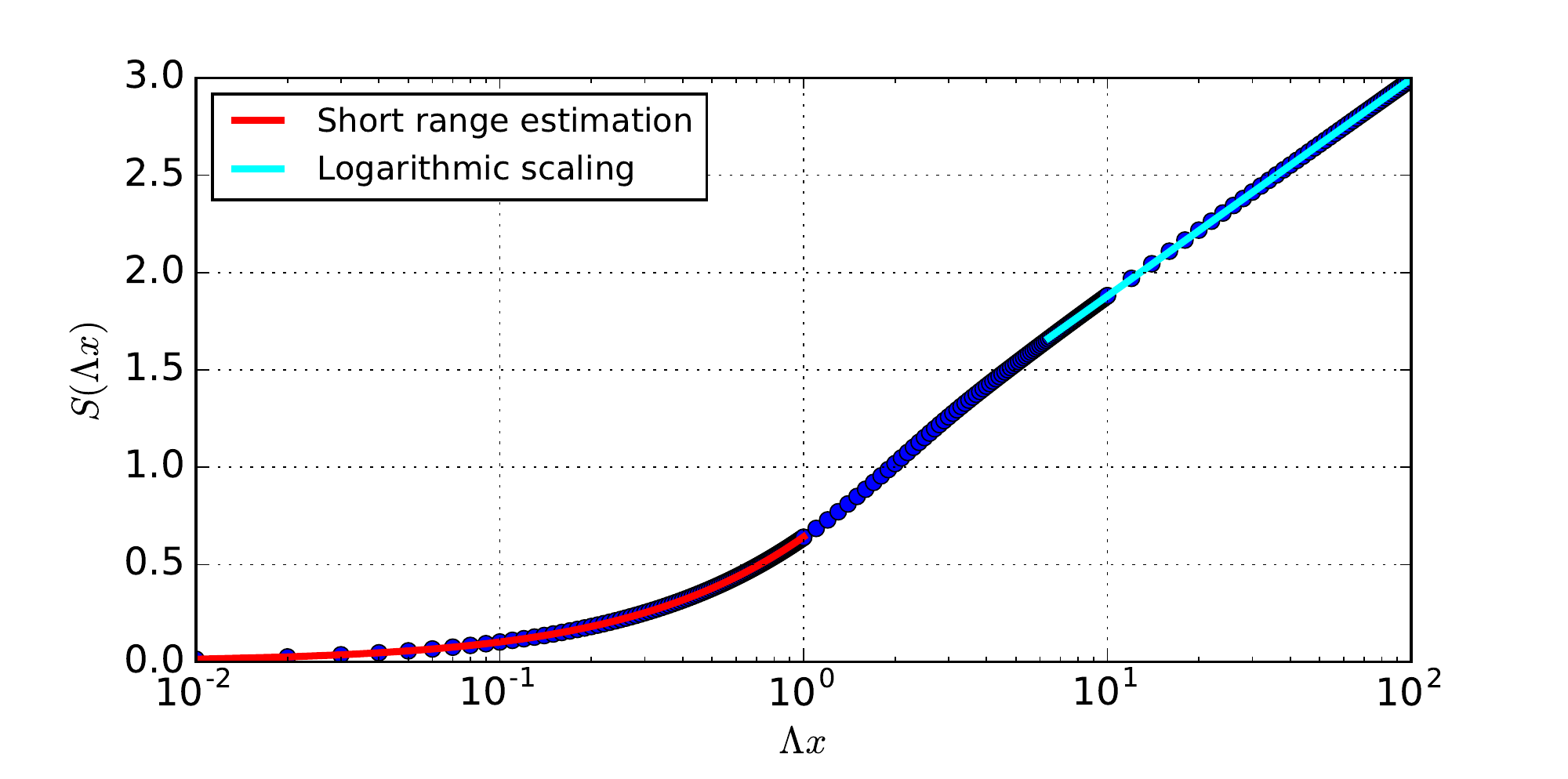}
			\caption{ Entanglement entropy profile obtained for $\Lambda a=0.01$ ($\Lambda x<10$) and $\Lambda a=0.1$ ($\Lambda x>10$). We superpose the short distance estimation and the fit to logarithmic scaling at distances much larger than the cutoff, which provides a value of the central charge $c\approx 1.003$.}
			\label{ent1dfermions}
		\end{figure}

\begin{figure}
			\centering
			\includegraphics[width=0.9\linewidth]{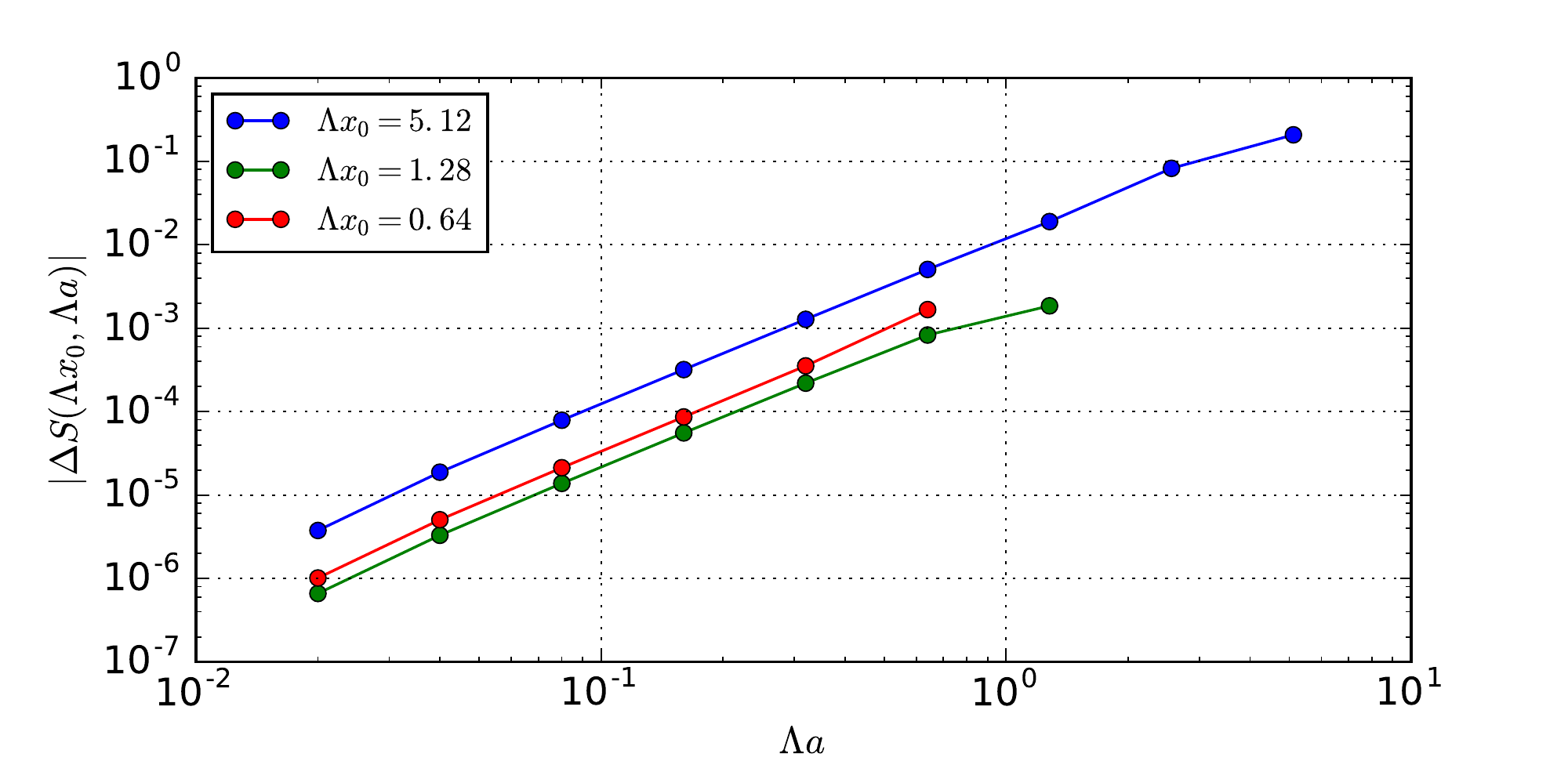}
			\caption{Plot of the difference $|S(\Lambda x_0,\Lambda a)-S(\Lambda x_0,\Lambda a = 0.01)|$ that shows the convergence of this particular value of entropy upon iterative fine-graining of the sampling parameter. The plotted difference goes to zero approximately quadratically with $\Lambda a$.}
			\label{conv1dfermions}
		\end{figure}

Once more, we perform the numerical computation of entanglement entropy in cMERA by sampling the correlators with a certain lattice spacing $a$, which produces discrete versions of the corresponding continuum correlation operators, to which we apply the usual prescription (see Appendix \ref{apa}). Once more we find finite values of the entropy to which our results converge when $a\to 0$, hinting at the removal of short scale entanglement. Figure \ref{ent1dfermions} shows the results of the numerical computation of entanglement entropy. As expected, two differentiated regimes are visible. For small intervals $x \ll 1/\Lambda$, the entanglement entropy $S(x)$ is seen to vanish as $x\to 0$, with the numerics matching an analytical estimation derived in Appendix \ref{apb}. For large intervals $a \gg 1/\Lambda$, 
the expected logarithmic scaling of $S(x)$ is recovered. The value obtained for the central charge from fitting the curve is very close to our expectation,
\begin{equation}
c\approx 1.003\approx 1.
\end{equation}
Importantly, we observe, as for the bosonic case, convergence of the entropy upon fine graining of the underlying lattice that we use to compute it, see Figure \ref{conv1dfermions}.
		
\subsection{Free massless fermion in 2+1 dimensions}
We use the two-dimensional representation of the Dirac $\gamma$ matrices given by:
\begin{equation} 
\gamma^0=\left(\begin{array}{cc}
1 & 0\\0 & -1
\end{array} \right),\qquad\gamma^1=\left(\begin{array}{cc}
0 & -i\\-i & 0
\end{array} \right),\qquad\gamma^2=\left(\begin{array}{cc}
0 & -1\\1 & 0
\end{array} \right).
\end{equation}
The unitary defining the cMERA belongs now to SU(2):
\begin{equation}
\left( \begin{array}{c}
\tilde\psi_1(k)\\
\tilde\psi_2(k)
\end{array}\right)=\left( \begin{array}{cc}
\cos\left(\theta(k)\right) & -ie^{-i\varphi_{\vec{k}}}\sin\left(\theta(k)\right)\\
-ie^{i\varphi_{\vec{k}}}\sin\left(\theta(k)\right) & \cos\left(\theta(k)\right)
\end{array}\right) \left( \begin{array}{c}
\psi_1(\vec k)\\
\psi_2(\vec k)
\end{array}\right)
\end{equation}
where $\varphi_{\vec{k}}$ is the angle between the momentum vector and the $x$-axis, such that
\begin{equation}
|\vec k|e^{i\varphi_{\vec k}}=k_x+ik_y.
\end{equation}

\subsubsection*{Two-point correlation functions}
We have the following two-point correlation functions in momentum space:
\begin{align}
&\la\psi^\dagger_1(\vec k)\psi^{\phantom{\dagger}}_1(\vec q)\ra=\sin^2(\theta(k))\delta(\vec k-\vec q)\label{fer21}\\
&\la\psi^\dagger_1(\vec k)\psi^{\phantom{\dagger}}_2(\vec q)\ra=-ie^{i\varphi_{\vec k}}\dfrac{\sin(2\theta( k))}{2}\delta(\vec k-\vec q)=\overline{\la\psi^\dagger_2(\vec k)\psi^{\phantom{\dagger}}_1(\vec q)\ra}\label{fer22}\\
&\la\psi^\dagger_2(\vec k)\psi^{\phantom{\dagger}}_2(\vec q)\ra=\delta(\vec k-\vec q)-\la\psi^\dagger_1(\vec k)\psi^{\phantom{\dagger}}_1(\vec q)\ra\label{fer23}
\end{align}
and in position space
\begin{align}
\la\psi^\dagger_1(\vec x)\psi^{\phantom{\dagger}}_1(\vec y)\ra&=\int_0^\infty{\dfrac{k\:dk}{2\pi}\sin^2(\theta(k))J_0(k|\vec x-\vec y|)}=f(|\vec x-\vec y|)\\
\la\psi^\dagger_1(\vec x)\psi^{\phantom{\dagger}}_2(\vec y)\ra&=
\dfrac{i}{8\pi^2}\int_0^\infty{k\:dk\sin(2\theta(k))\int_0^{2\pi}{d\phi\: e^{-i(\phi+k|\vec x-\vec y|\cos\phi)}}}e^{i\phi_{\vec x-\vec y}}\nonumber\\&=g(|\vec x - \vec y|)e^{i\phi_{\vec x-\vec y}}=\overline{\la\psi^\dagger_2(\vec k)\psi_1(\vec q)\ra}\\
\la\psi^\dagger_2(\vec x)\psi^{\phantom{\dagger}}_2(\vec y)\ra&=\delta(\vec x-\vec y)-\la\psi^\dagger_1(\vec x)\psi^{\phantom{\dagger}}_1(\vec y)\ra.
\end{align}
Note that the correlator between the same fermionic species depends only on the distance between the points, while the correlator between different species carries a phase related to the orientation of the vector $\vec x-\vec y$.
\begin{figure}
	\centering
	\includegraphics[width=0.8\linewidth]{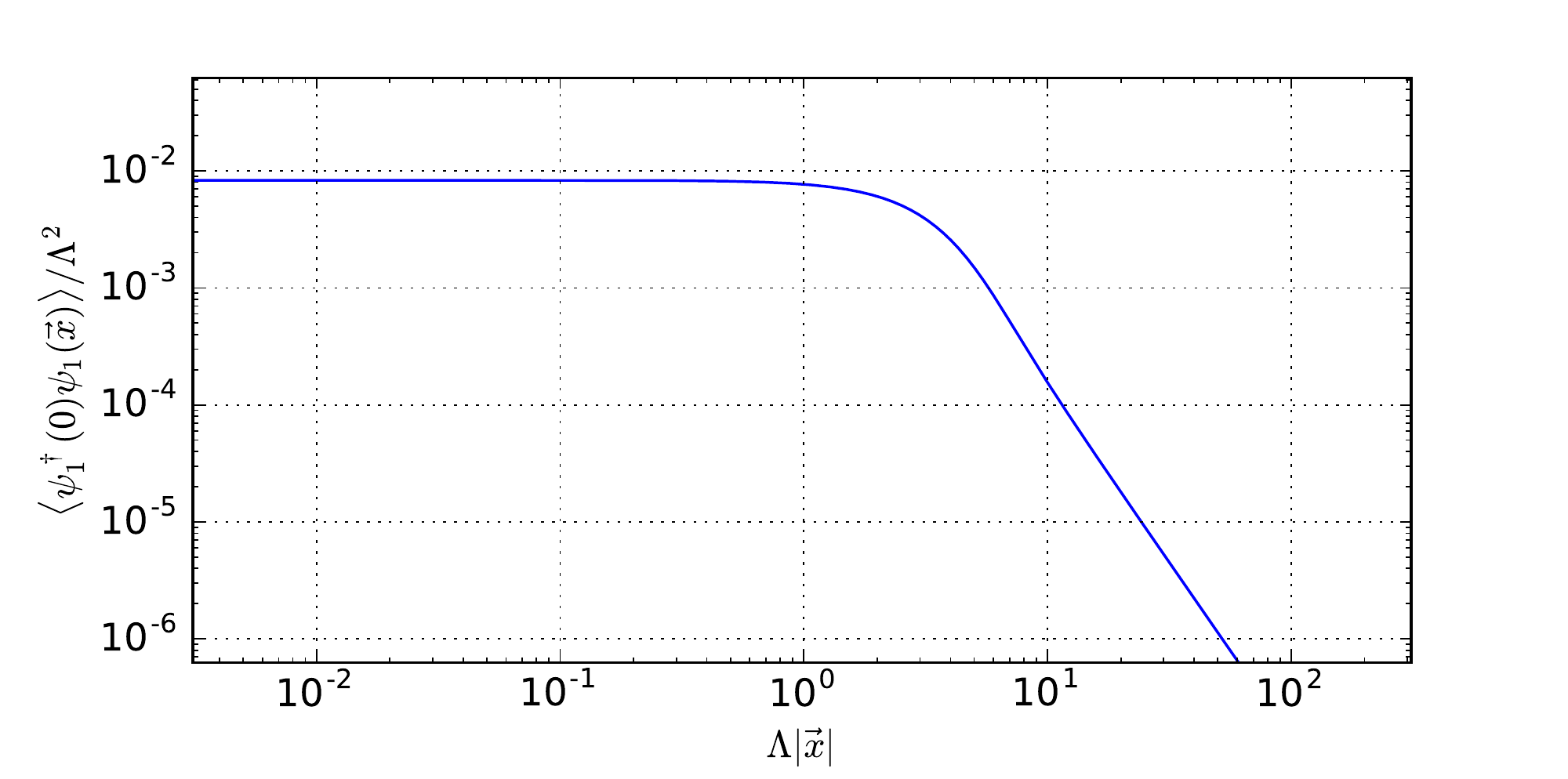}
	\includegraphics[width=0.8\linewidth]{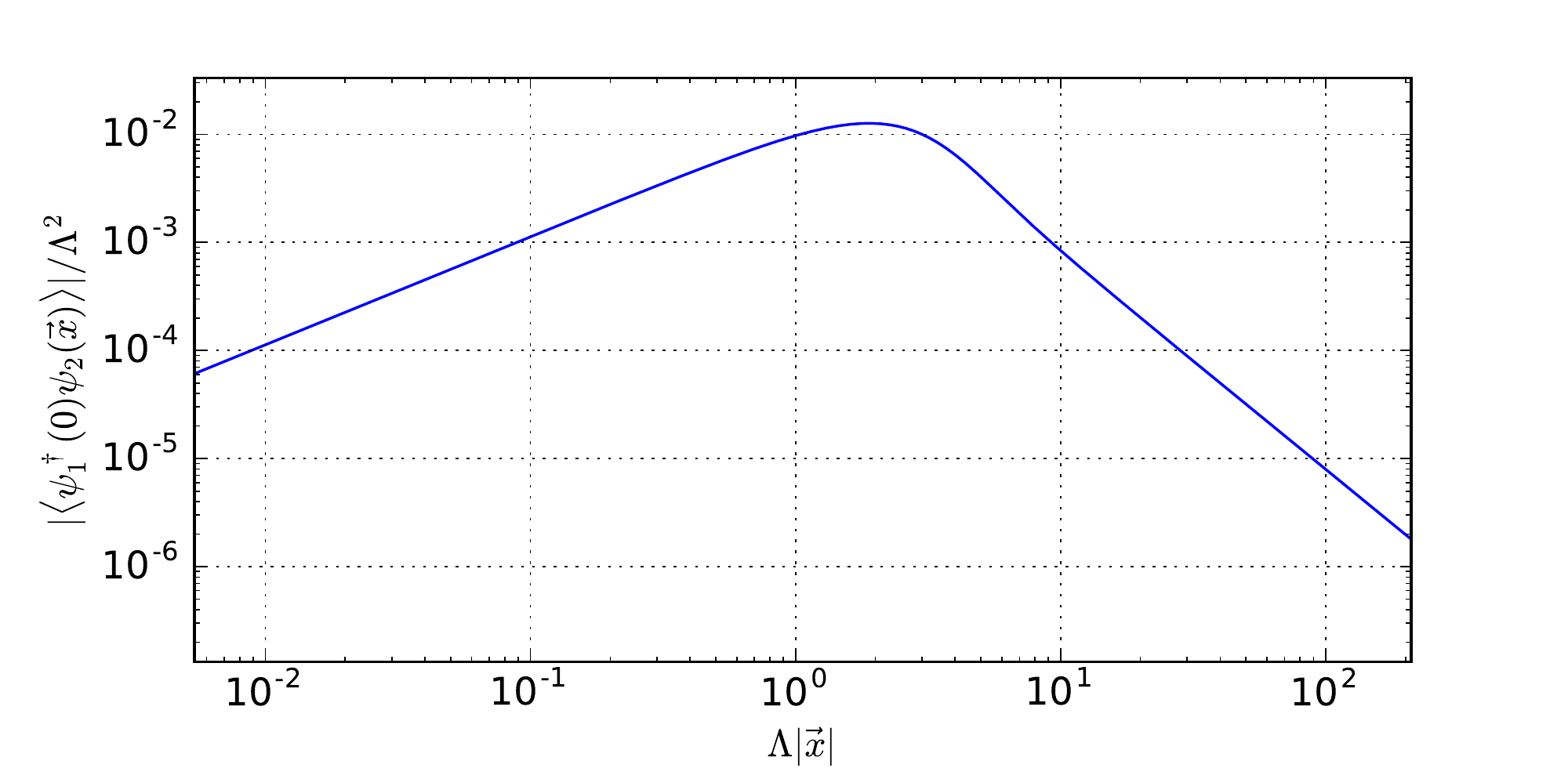}
	\caption{Correlation  functions $\la\psi_1^\dagger(0)\psi_1(\vec x)\ra$ (top) $\la\psi_1^\dagger(0)\psi_2(\vec x)\ra$ (bottom, in absolute value) and computed for the 2+1 dimensional fermionic cMERA with $j=0$.}
	\label{corr2dfermions}
\end{figure}

The long distance decay properties of the two-point functions can once again be deduced from their momentum space representation (\ref{fer21}~--~\ref{fer23}) by means of the methods reviewed in Appendix \ref{asymfou}. For the single-species correlators, the cMERA again displays a spurious power-law decay $|\vec x-\vec y|^{-2j-3}$, which was not present in the CFT. This is a consequence of the behaviour of the function $\sin^2(\theta(\vec k))$ around $\vec k=0$: it goes as $|\vec k|^{2j+1}$. In the case of the two-species correlator, however, we infer a leading decay given by $|\vec x-\vec y|^{-2}$, due to the presence of the phase factor in $\eqref{fer22}$, independently of $j$. This is also the case for the actual CFT, since the scaling dimension of the field $\psi$ is 1 in 2+1 dimensions. The next-to-leading term gives the leading order cMERA correction to the CFT correlators, and goes like:
	\begin{equation}
|\la\psi_1^\dagger(\vec x)\psi^{\phantom{\dagger}}_2(\vec y)\ra_{\text{cMERA}}-\la\psi_1^\dagger(\vec x)\psi^{\phantom{\dagger}}_2(\vec y)\ra_{\text{CFT}}|\sim|x-y|^{-(4j+4)}.
	\end{equation}

Figure \ref{corr2dfermions} shows how correlations behave for this cMERA, again for $j=0$. The usual two regimes are observed: at distances smaller that $1/\Lambda$, the single-species correlator is practically constant, and the two-species correlator grows linearly (as a power law $|\vec x-\vec y|^{p}$ with $p\approx 0.9994\approx 1$) with distance. Once the cutoff length scale is surpassed, the single-species correlator decays with exponent $p\approx -3.008\approx-3$, while the two-species correlator decays with exponent $p\approx -2.005\approx-2$, which is in accordance with the argumentation at the beginning of this paragraph.

\subsubsection*{Entanglement entropy}
\begin{figure}
	\centering
	\includegraphics[width=0.9\linewidth]{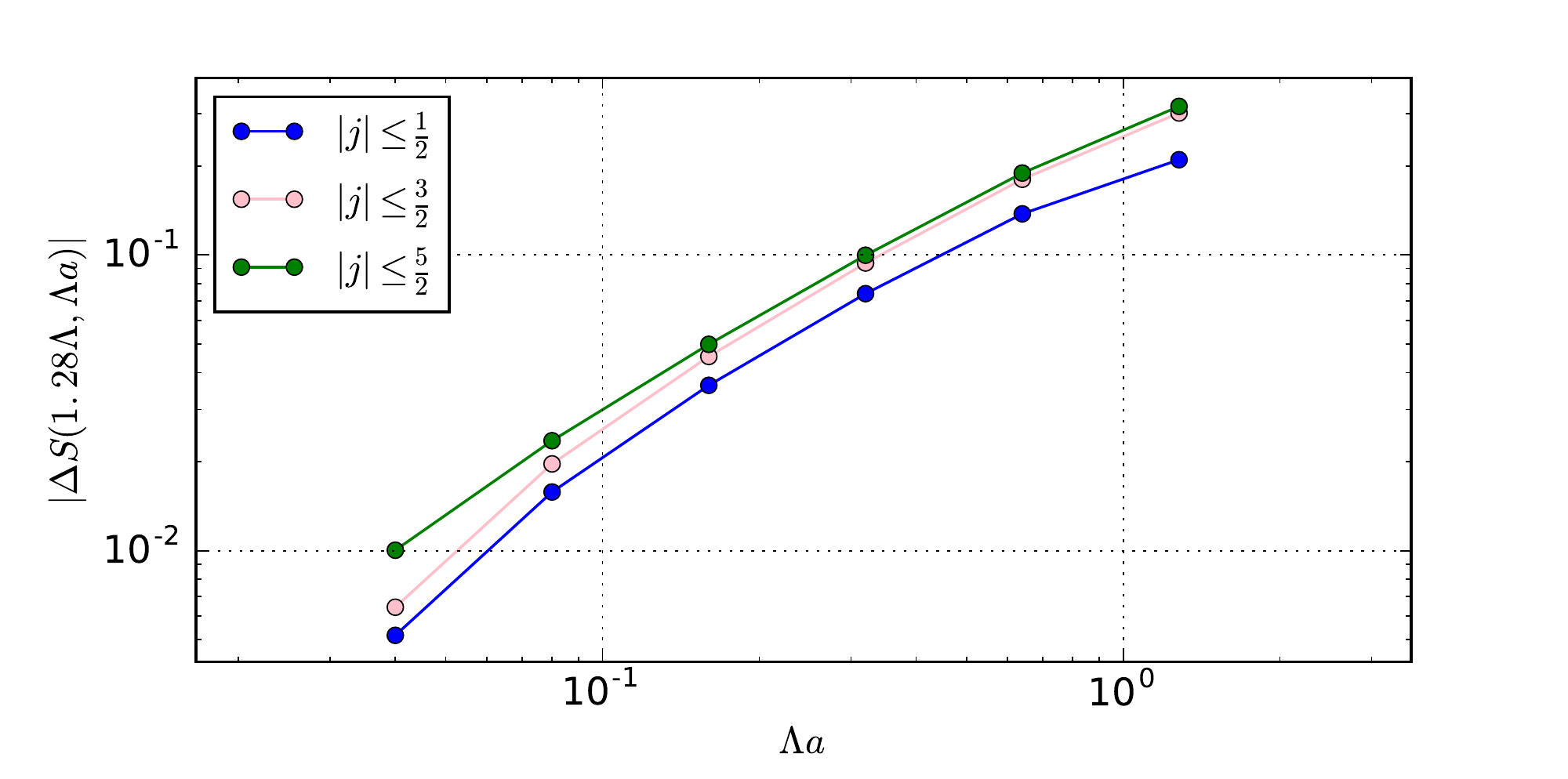}
	\caption{Plot of the difference $|S(1.28\Lambda,\Lambda a)-S(1.28\Lambda,\Lambda a = 0.02)|$, for different maximum values of $|j|$ that shows the convergence of this particular value of entropy upon iterative fine-graining of the sampling parameter $a$. Note how truncating at higher values of $|j|$ does not affect very noticeably the rate of convergence in $a$.}
	\label{fig:ent2dfermionsconvergence}
\end{figure}

As in the 2+1 dimensional bosonic case discussed earlier, we will study the scaling of entanglement entropy $S(x)$ by taking discs of increasing radii $x$ and tracing out the rest of the system. We do so by again changing to polar coordinates and taking into account only the modes with small angular momentum, which provide the main contributions. The details of the computation are presented in Appendix \ref{apc}. 
		\begin{figure}
			\centering
			\includegraphics[width=0.9\linewidth]{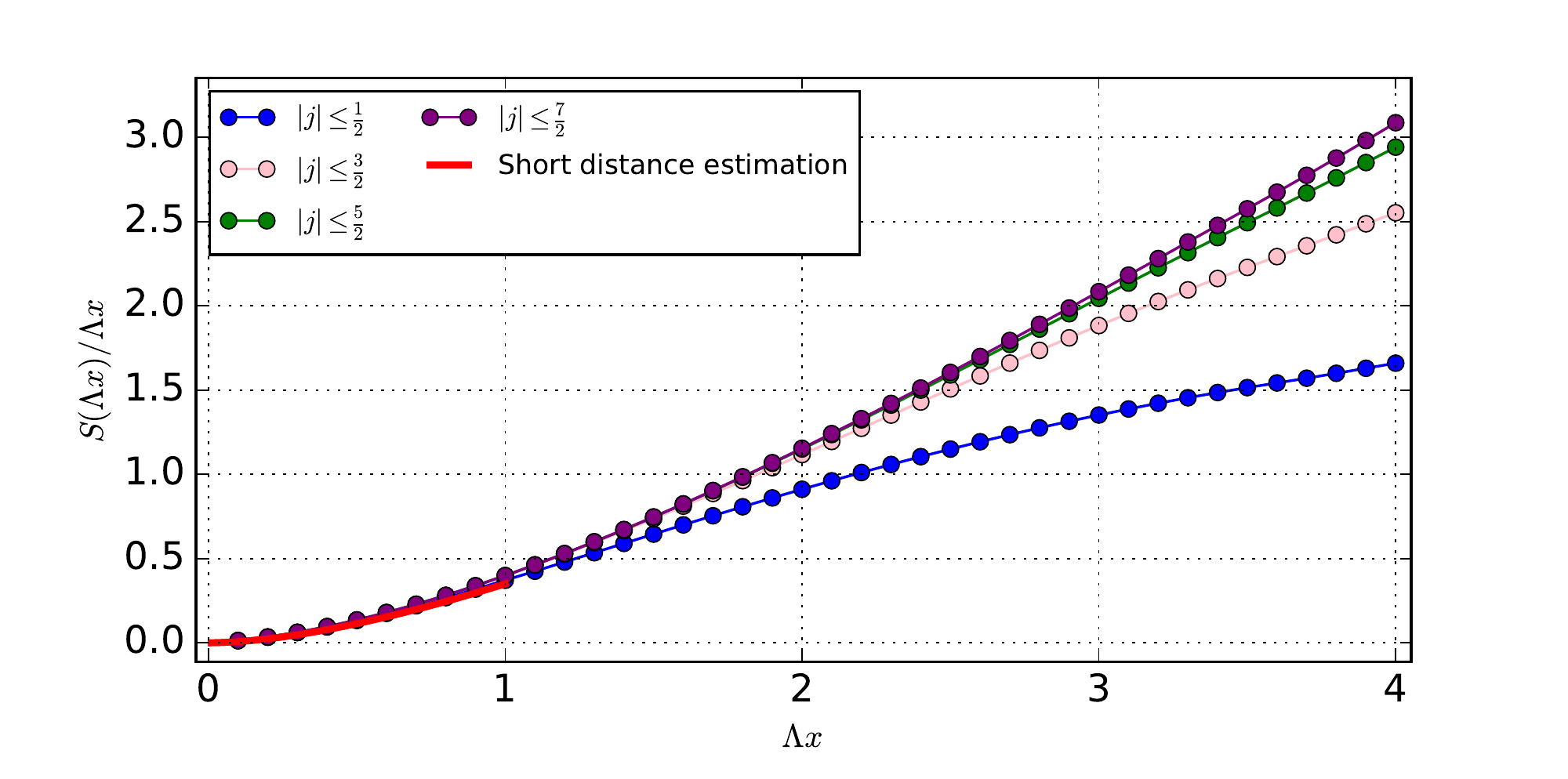}
			\caption{Entanglement entropy profile obtained for $\Lambda a=0.01$ ($ x<4/\Lambda$). Convergence upon increase of the maximum value of $|j|$ is observed. We lack numerical data to comment on how reliably the long distance behaviour reproduces an area law.}
			\label{ent2dfermions}
		\end{figure}
Figures \ref{ent2dfermions} and \ref{fig:ent2dfermionsconvergence} show that the main features we have been observing in this paper prevail in the 2d fermionic case: the entropy converges to a finite value upon fine-graining of the sampling of the correlators. However, the computations became too heavy before we could assert with enough confidence the presence of two differentiated regimes, or the recovery of the area law (note that the onset of the area law in Figure \ref{ent1dfermions} can only be clearly appreciated past $x \sim 3/\Lambda$, where we stop having converged data in the 2d case). Nevertheless, we can still give a good analytical estimation of the scaling of the entropy (see Appendix \ref{apb}) for small radii $x\ll 1/\Lambda$.

	\section{Conclusions and discussion}
		\label{sec5}
		In this paper we have studied the structure of correlations and entanglement in Gaussian cMERA states optimized to approximate the ground states of free particle CFTs.
		 Our results suggest that, in line with what was argued by the proponents of this ansatz in Ref. \cite{Haegeman2011}, cMERA states come endowed with a characteristic length scale $1/\Lambda$ that separates two very different regimes. At length scales larger than $1/\Lambda$, the correlation structure, as observed via two-point correlation functions and entanglement entropy scaling, reproduces the features of the target CFT ground state. At length scales smaller than $1/\Lambda$, however, the behaviour changes in a way we have characterized, and becomes more akin to that of a product state. This has observable consequences such as the removal of the UV divergence of entanglement entropy, which instead acquires a finite value. As a result, we refer to cMERA as having a UV cutoff in its entanglement.

		This matches the intuition we obtain by revisiting Figure \ref{MERA}, which compares the MERA and cMERA schemes. In MERA, the disentanglers allow for the introduction in the initial state of entanglement at scales no shorter that the spacing that separates two lattice sites. This is an obvious statement right from the onset, since there are no degrees of freedom to entangle between these two sites. In the QFT setting, there exists instead a continuum of degrees of freedom between any two given points, but the entangler $K$ is designed in such a way that it cannot entangle at distances shorter than $1/\Lambda$. Thus the result of an evolution by $L+K$, namely the cMERA state $\ket{\Psi^{\Lambda}}$ behaves in such a different manner as the target state $\ket{\Psi}$ when probed at length scales shorter than $1/\Lambda$.

We have focused on (free particle) CFTs because of their scale invariance, which significantly simplifies the analysis. However, most of our conclusions extend straightforwardly to (free particle) massive QFTs, provided that the length scale $\xi\sim 1/m$ due to the mass $m\not = 0$ is larger than the cutoff distance $1/\Lambda$. In fact, there is nothing that suggests that this should not apply also to non-Gaussian cMERA states constructed for interacting QFTs. The development of such non-Gaussian MERA states could bring prospects of providing a systematic UV regularization scheme for quantum field theories.

	\appendix

		\section{Asymptotics of two-point functions}
		\label{asymfou}
		In this Appendix we review the analytical determination of the asymptotic decay of two-point functions at long distances. This is done by arguments of asymptotic analysis which we expose in a self-contained manner. All the momentum space two-point functions we find in this work are of the form
		\begin{equation}
		\la \mc O(\vec k) \tilde{\mc O}(\vec q) \ra = f(\vec k) \delta(\vec k\pm \vec q),
		\end{equation}
	 the variable sign being + for bosonic theories and -- for fermionic theories. The correlator in position space is then given by the inverse Fourier transform of $f(\vec k)$, up to a proportionality constant:
		\begin{equation}
		\la \mc O(\vec 0) \tilde{\mc O}(\vec x) \ra = (2\pi)^{-d/2} \mc F^{-1}[f](\vec x).
		\end{equation}
		 where $d$ is the spatial dimension. We proceed now to argue how the asymptotic properties of this correlator for large $|\vec x|$ can be inferred from the knowledge of $f(\vec k)$.

		\subsection{1+1 dimensions}
		Let us first consider the case with one spatial dimension. Assume $f(k)$ is integrable. The Riemann-Lebesgue lemma then states that $\mc F^{-1}[f](x)$ has to decay to zero at long distances:
		\begin{equation}
			f \in L^1(\R)\implies \mc F^{-1}[f](x)=\int{\dfrac{dk}{\sqrt{2\pi}}e^{ikx}f(k)}\to 0\qquad x\to\pm\infty.
		\end{equation}
		Imposing further conditions on $f(k)$ allows us to be more precise in the characterization of this long distance decay.
		For example, if we assume that $f'(k)$ exists and is also in $L_1(\R)$. Then applying the derivative rule of the Fourier transform gives
		\begin{align}
			\mc F^{-1}[f](x)=\dfrac{i}{x}\mc F^{-1}[f'](x).
			\label{derrule}
		\end{align}
		But now, by the same Riemann-Lebesgue lemma, $f'(k)\in L^1(\R)$ implies that $\mc F^{-1}[f'](x)$ also vanishes in the limit $|x|\to\infty$. Thus, $\mc F^{-1}[f]$ itself goes to zero faster than $|x|^{-1}$ when $|x|\to \infty$.
		
		Alternatively, let us consider what would happen if $f(k)$ were differentiable with integrable derivative except at a number of jump discontinuities of size $\Delta_i$ at points $\{k_i\}$:
		\begin{equation}
			\lim_{\varepsilon\to 0^+}f(k_i+\varepsilon)-f(k_i-\varepsilon)=\Delta_i,
		\end{equation}
		so that $f'(k)=h(k)+\sum_i{\Delta_i\delta(k-k_i)}$ with $h(k)\in L_1(\R)$. Then we have to rewrite \ref{derrule} as:
		\begin{equation}
		\mc F^{-1}[f](x)=\dfrac{i}{x}\mc F^{-1}[f'](x)+\dfrac{i}{x}\sum_i{\dfrac{\Delta_ie^{ik_ix}}{\sqrt{2\pi}}},
		\end{equation}
		and the new term on the right hand side becomes the first term in the asymptotic series expansion of $F^{-1}[f](x)$. It will dominate at long distances, decaying as a power law $|x|^{-1}$ together with a certain oscillation\footnote{For instance, this kind of oscillations in the correlators of a fermionic system carry information of the position of the Fermi surface.} dependent on the values of the $k_i$.
		
		If we are in the situation where $f'(k)\in L^1(\R)$, this same argument above can be applied to $f''(k)$, and iteratively to higher derivatives. If $f(k), f'(k),\ldots,f^{(n-1)}(k)$ all exist as continuous integrable functions, but $f^{(n)}(k)$ presents jump discontinuities, we will find the leading order decay of $F^{-1}[f](x)$ at long distances to be $|x|^{-(n+1)}$.
			
		\subsection{2+1 dimensions}
			
		In higher dimensions, $f(\vec k)$ can display a higher variety of features that translate into asymptotic properties of $\mc F^{-1}[f](\vec x)$, and hence of the position space correlator. In this work we encounter two different situations, depending on the behaviour of $f(\vec k)$ around the origin, which is the only point where it is not infinitely differentiable. We review both cases in what follows. Our exposition in this section partially draws from \cite{lamoureux}.
		\subsubsection*{First case: $f(\vec k) \sim |\vec k|^{2n-1},\; n\geq 0$ as $|\vec k|\to 0$}
		The $\la\phi\phi\ra$ and $\la\pi\pi\ra$ bosonic correlators, and the single-species fermionic correlators all belong to this first case, since close to the origin we have 
		\[f(\vec k)\sim |\vec k|^{-1},\qquad f(\vec k)\sim |\vec k|,\qquad\text{and}\qquad f(\vec k)\sim |\vec k|^{2j+1}\]
		respectively for each of them. In general, $f(\vec k) \sim |\vec k|^{2n-1}$ is associated with a $|\vec x|^{-2n-1}$ decay. This fits in well with the picture we obtained from the 1+1-dimensional case, since a higher $n$ means higher order for the first discontinuous derivative of $f(\vec k)$ at the origin.
		
		Let us see it for the particular case of $n=0$, when $f(\vec k)$ presents a $1/|\vec k|$ singularity at the origin. We define a new function
		\begin{equation}
		g(\vec k)=\dfrac{e^{-|\vec k|^2}}{|\vec k|}\in L_1(\R^2),
		\end{equation}
		which we use to subtract the singularity\footnote{Of course, if needed we could multiply $g$ by an appropriate constant omitted here.}:
		\begin{equation}
		f(\vec k) = g(\vec k) + h(\vec k),
		\end{equation}
		so that the components of $\vec\nabla h(\vec k)$ are integrable (though they might be discontinuous). Then we have 
		\begin{equation}
		\mc F^{-1}[f](\vec x)=\mc F^{-1}[g](\vec x)+\mc F^{-1}[h](\vec x),
		\end{equation}
		and we can deal with each of the terms individually. For $h$ we have
		\begin{equation}
		|F^{-1}[h](\vec x)|=\dfrac{1}{|\vec x|}|\mc F^{-1}[\vec \nabla h](\vec x)|,
		\label{hache}
		\end{equation}
		and the Riemann-Lebesgue lemma again forces the right hand side to decay faster than $|\vec x|^{-1}$. However, for $g(\vec k)$ we have
		\begin{align}
		g(\vec k)=e^{-|\vec k|^2}\cdot\dfrac{1}{|\vec k|}\implies\mc F^{-1}[g](\vec x) &= \mc F^{-1}\left[e^{-|\vec k|^2}\right](\vec x) * \mc F^{-1}\left[\dfrac{1}{|\vec k|}\right](\vec x)\propto\\&\propto\int_{\R^2}{d^2y\;\dfrac{e^{-|\vec y|^2/4}}{|\vec x - \vec y|}},
		\end{align}
		where $*$ denotes the convolution product. It is easy to see that the result of the convolution decays as $|\vec x|^{-1}$ by bounding it above and below. We have
		\begin{align}
		\int_{\R^2}{d^2y\;\dfrac{e^{-|\vec y|^2/4}}{|\vec x - \vec y|}}&=\int_{B\left(\vec x, \frac{|\vec x|}{2}\right)}{d^2y\;\dfrac{e^{-|\vec y|^2/4}}{|\vec x - \vec y|}}~~+~~\int_{\R^2\backslash B\left(\vec x, \frac{|\vec x|}{2}\right)}{d^2y\;\dfrac{e^{-|\vec y|^2/4}}{|\vec x - \vec y|}}\leq\nonumber\\[10pt]&\leq  e^{-|\vec x|^2/16}\int_{B\left(\vec x, \frac{|\vec x|}{2}\right)}{d^2y\;\dfrac{1}{|\vec x - \vec y|}}~~+~~\dfrac{2}{|\vec x|}\int_{\R^2}{d^2y\;e^{-|\vec y|^2/4}}\leq\nonumber\\[10pt]&\leq \pi |\vec x|e^{-|\vec x|^2/16}+\dfrac{8\pi}{|\vec x|},
		\end{align}
		where $B(\vec a, r)$ is the ball of radius $r$ centered at $\vec a$. Equally,
		\begin{align}
		\int_{\R^2}{d^2y\;\dfrac{e^{-|\vec y|^2/4}}{|\vec x - \vec y|}}&\geq\int_{B\left(-\frac{\vec x}{|\vec x|}, 1\right)}{d^2y\;\dfrac{e^{-|\vec y|^2/4}}{|\vec x - \vec y|}}\geq\dfrac{e^{-1}}{|\vec x|}\int_{B\left(-\frac{\vec x}{|\vec x|}, 1\right)}{d^2y}=\dfrac{\pi}{e}\dfrac{1}{|\vec x|}.
		\end{align}
		Thus, we have proved that the leading order of decay of $\mc \mc F^{-1}[f](\vec x)$ is $|\vec x|^{-1}$ as claimed.
		
		For higher values of $n$ we can proceed by induction. If $f(\vec k)\sim |\vec k|^{2n-1}$ we subtract the function:
		\begin{equation}
		g(\vec k)=e^{-\vec k^2}|\vec k|^{2n-1},
		\end{equation}
		and expect the derivatives of $h(\vec k) = f(\vec k)-g(\vec k)$ of order up to at least $2n+1$ to be in $L^1(\R^2)$. This assures that $\mc F^{-1}[h](\vec x)$ decays faster than $|\vec x|^{-(2n+1)}$ by iterating an argument like the one from Eq. \ref{hache}.
		Now, by the induction hypothesis
		\begin{align}
		\mc F^{-1}[g](\vec x)&=-\dfrac{\mc F^{-1}[\Delta g](\vec x)}{|\vec x|^2}=\nonumber\\&=-\dfrac{1}{|\vec x|^2}\mc F^{-1}\left[e^{-|\vec k|^2} \left((1-2 n)^2 |\vec k|^{2 n-3}-8 n  |\vec k|^{2 n-1}+4  |\vec k|^{2 n+1}\right)\right](\vec x)=\nonumber\\&=O\left(|\vec x|^{-(2n+1)}\right) .
		\end{align}
		
		\subsubsection*{Second case: $f(\vec k) \sim k^{2n}e^{i\phi_{\vec k}},\;n\geq 0$ as $|\vec k|\to 0$}
		
		 We denote by $\phi_{\vec k}$ the angle between $\vec k$ and the horizontal axis of the plane. This is the case, with $n=0$, for the fermionic two-species correlator, and is associated to a $|\vec x|^{-(2n+2)}$ decay. To see this, we use a very similar strategy as in the first case. Consider
		 \begin{equation}
		 g(\vec k)=e^{i\phi_{\vec k}}k^{2n}e^{-|\vec k|^2},
		 \end{equation}
		and write
		\begin{equation}
		f(\vec k)=g(\vec k)+h(\vec k)\qquad\text{  with  }\qquad \Delta ^{n+1}h\in L^1(\R^2).
		\end{equation}
		The condition on the $(n+1)$-th power of the Laplacian of $h$ guarantees that
		\begin{equation}
		\mc F^{-1}[h](\vec x)=(-1)^{n+1}\dfrac{\mc F^{-1}[\Delta^{n+1} h](\vec x)}{|\vec x|^{2n+2}}
		\end{equation}
		decays faster than $|\vec x|^{-(2n+2)}$. However, for a radially symmetric function $G(\vec k)$ in two dimensions it holds
		\begin{align}
		\mathcal{F}^{-1}[e^{i\phi_k}G(\vec k)](\vec x)&=-i\partial_{|\vec x|}\mathcal{F}^{-1}\left[\dfrac{G(\vec k)}{|\vec k|}\right](\vec x),
		\end{align}
		provided that $G(k)/|\vec k|$ is integrable. Applying this expression for $g(\vec k)=e^{i\phi_{\vec k}}G(\vec k)$ already implies $\mathcal{F}^{-1}[g](\vec x)=O\left( |\vec x|^{-(2n+2)}\right)$, if we use the results from the first case above.

	\section{Correlation measures in Gaussian states}
	\label{apa}
	The tools that we used in the main text to probe the short distance properties of cMERA states are the two-point correlation functions and the entanglement entropy profile. Both are measures of the correlations between different degrees of freedom of the theory, and we can use them to study how these correlations depend on the scale at which we are looking.
	
	The two-point correlation functions of the theory are $\la\mc O(x)\mc O'(y)\ra$, where $\mc O(x),\mc O'(y)$ belong to the local algebra of operators at their respective locations. Their role in theories with quadratic Hamiltonians, such as the free theories we deal with in this paper is particularly important, since they encode all the information of the $N$-point functions for the ground state of the theory. Indeed, only the correlation matrix of the field modes is needed to completely specify the state. This is due to the state being Gaussian, which implies that it satisfies Wick's theorem: given operators $\mc O_1,\ldots,\mc O_N$ that are linear in the field modes, we have
	\begin{equation}
	\la\mc O_1\ldots\mc O_N \ra=\sum_{\sigma=(i_1,\ldots,i_N)}{s_\sigma\la \mc O_{i_1}\mc O_{i_2}\ra\ldots\la\mc O_{i_{N-1}}\mc O_{i_N}\ra}
	\end{equation}
	where the sum runs over all possible pairings of the operators, and $s_\sigma$ is a sign that accounts for the commutation/anticommutation relations of the modes.
	
	Our second witness for correlations is entanglement entropy. 
	Given a certain spatial region $\mc R$, we may obtain its associated density matrix $\rho_{\mc R}$ by tracing out the degrees of freedom outside $\mc R$. If we are then able to find a series of uncorrelated modes supported on $\mc R$ via an appropriate change of variables (canonical transformation), the density matrix will factor into the tensor product of the density matrices associated to said modes. These will each be in a thermal state, and the total entropy of the state can be computed as a sum over contributions from each mode.
	\begin{equation}
	\rho_{\mc R}=\bigotimes_{i}{\rho_i}\implies S(\rho)=\sum_i{S(\rho_i)}.
	\label{fact}
	\end{equation}
	
	The procedure needed to find these modes is different but analogous depending on whether we are speaking of bosons or fermions \cite{Casini2009}. In Appendix \ref{apb} we perform some estimations while still in the continuum, but in the main text we discretize the correlation matrix by sampling of the two-point functions. We can then interpret it as the (approximate) correlation matrix for a discrete, finite set of bosonic/fermionic modes, and carry on from there, with the advantage that numerical computation then becomes available as a tool to produce the final results. We therefore proceed to review the entanglement entropy computation techniques we employ in Gaussian states of finitely many modes.
	
	\subsection{Bosonic theories}
	
	By discretizing a bosonic theory we obtain an algebra of operators $\{\phi_i,\pi_i\}$ that satisfy the canonical commutation relations (CCR):
	\begin{equation}
	[\phi_i,\pi_j]=i\delta_{i,j}\,,~~~~~~[\phi_i,\phi_j]=[\pi_i,\pi_j]=0.
	\end{equation}
	Linear transformations in this algebra that preserve the CCR form a group and are called canonical transformations. They map bosonic modes into bosonic modes. The group of canonical transformations for $N$ bosonic modes is the symplectic group $\text{Sp}(2N,\C)$, which by definition is the subgroup of $\text{GL}(2N,\C)$ whose elements $M$ satisfy
	\begin{equation}
		M\left( \begin{array}{cc}
			0&\mathds{1}_N\\
			-\mathds{1}_N&0
		\end{array}\right)M^T=\left( \begin{array}{cc}
		0&\mathds{1}_N\\
		-\mathds{1}_N&0
	\end{array}\right).
	\end{equation}
	(Notice that this is precisely the condition that the CCR are preserved when the map $M$ is applied to the column vector of modes $(\phi_1,\ldots,\phi_N,\pi_1,\ldots,\pi_N)$.)
	
	A Gaussian state, as stated above, is completely characterized by its Hermitian, positive definite correlation matrix:
	\begin{equation}
		C^{\mc O\mc O'}_{ij}=\la\mc O_i\mc O'_j\ra\text{,    with  }\mc O,\mc O'\in\{\phi,\pi\},\;i,j=1,\ldots N
	\end{equation}
	Given any such matrix, there exists a procedure, called \textit{symplectic diagonalization}, by which we can find a symplectic transformation that maps our initial set of bosonic modes to a set of uncorrelated modes $\{\tilde{\phi}_i,\tilde\pi_i\}$, so that the correlation matrix decomposes as a direct sum \cite{Eisert2010}:
	\[C=\bigoplus_{i=1}^n{\left( \begin{array}{cc}
		\langle\tilde\phi_i\tilde\phi_i\rangle&\langle\tilde\phi_i\tilde\pi_i\rangle\\
		\langle\tilde\pi_i\tilde\phi_i\rangle&\langle\tilde\pi_i\tilde\pi_i\rangle
		\end{array}\right)}.\]
	Consequently the density matrix factorizes as in \eqref{fact}. Each of the uncorrelated modes will be in a thermal state of the following form:
	\begin{equation}
	\rho_i=(1-\zeta_i)\sum_{n=0}^\infty{\zeta_i^n\ket{n}\hspace{-0.5mm}\bra{n}}\implies S(\rho_i)=-\dfrac{\zeta_i\log_2{\zeta_i}}{1-\zeta_i}-\log_2(1-\zeta_i)
	\end{equation}
	where $\zeta_i\in[0,1)$ can be written in terms of the eigenvalues $\lambda_i$ of the matrix
	\begin{equation}
	K_{ij}=C^{\phi\phi}_{ik} C^{\pi\pi}_{kj},
	\label{opent}
	\end{equation}
	that is, the product of the $\phi\phi$ and the $\pi\pi$ submatrices of the correlation matrix\footnote{The set $\{\sqrt{\lambda_i}\}$ is usually referred to as the \textit{symplectic eigenvalues} of the correlation matrix $C^{\mc O\mc O'}_{ij}$.}. Indeed, we have
	\begin{equation}
	\zeta_i=\dfrac{2\sqrt{\lambda_i} -1}{2\sqrt{\lambda_i}+1}.
	\label{eigen}
	\end{equation}
	This way we can easily compute the entanglement entropy of a spatial region from its correlation matrix.
	
	\subsection{Fermionic theories}
	If we start with a fermionic theory and discretize it we arrive at a set of fermionic modes $\psi_i,\psi_i^\dagger$ which in turn satisfy the canonical anticommutation relations (CAR):
	\begin{equation}
	\{\psi_i^{\phantom{\dagger}},\psi^\dagger_j\}=\delta_{i,j}\,,~~~~~~\{\psi_i,\psi_j\}=\{\psi^\dagger_i,\psi^\dagger_j\}=0.
	\end{equation}
	The group of canonical transformations will this time be composed of those maps that preserve the CAR. It can be seen that for $N$ fermionic modes this group is isomorphic to $\text{O}(2N)$, the orthogonal group of dimension $2N$. For our purposes it will nonetheless be enough to consider the $U(N)$ subgroup given by the transformations:
	\begin{equation}
	\psi_i\longmapsto U_{ij}\psi_j
	\end{equation}
	where $U_{ij}$ is a unitary matrix. Note that these are the transformations that leave invariant the total particle number operator
	\begin{equation}
	\psi_1^\dagger\psi_1^{\phantom{\dagger}}+\ldots+\psi^\dagger_N\psi^{\phantom{\dagger}}_N.
	\end{equation}
	The correlation matrix that characterizes our Gaussian states will now be of the form:
	\begin{equation}
		C_{ij}=\la\mc \psi^\dagger_i\mc \psi^{\phantom{\dagger}}_j\ra\text{,  with  }i,j=1,\ldots N.
	\end{equation}
	And it can be proved that finding a canonical transformation that yields uncorrelated modes amounts to finding a unitary that diagonalizes this Hermitian matrix. The resulting modes will satisfy
	\begin{equation}
	\la\tilde\psi^\dagger_i\tilde\psi_j^{\phantom{\dagger}}\ra=\lambda_i\delta_{i,j}
	\end{equation}
	for $\lambda_i\in[0,1]$ the eigenvalues of the correlation matrix. Thus we can easily compute the entanglement entropy of the state given its correlation matrix:
	\begin{equation}
	S(\rho)=\sum_{i=1}^N{S(\lambda=\lambda_i)}=-\sum_{i=1}^N{\left[ \lambda_i\log\lambda_i+(1-\lambda_i)\log{(1-\lambda_i)}\right] }
	\end{equation}
	where $s(\lambda_i)$ is the entropy of the state of a single fermionic mode $\tilde\psi_i$ whose density matrix is 
	\begin{equation}
	\rho_i=(1-\lambda_i)\ket{0}\hspace{-0.5mm}\bra{0}+\lambda_i\ket{1}\hspace{-0.5mm}\bra{1}
	\end{equation}
	in the basis of eigenstates of the number operator $\tilde\psi^\dagger_i\tilde\psi^{\phantom{\dagger}}_i$.
	\section{Computation technicalities in 2+1 dimensions with rotational invariance}
	\label{apc}
	In the main text, and also in Appendix \ref{apb} we compute entanglement entropies for concentric discs centered at the origin, so that we can make use of the rotational invariance of the cMERA states. To do so, we perform a canonical transformation that reexpresses our fields in terms of the radial coordinate $r$ and an integer related to angular momentum \cite{Srednicki1993}. Here we specify how this is done for both bosonic and fermionic fields.
	\subsection{Bosonic theories}
	We define a new set of modes indexed by the radial coordinate $r$ and the integer $\ell$ that accounts for the angular component:
	\begin{align}
	&\phi_\ell(r)=\sqrt{\dfrac{r}{\pi}}\int_0^{2\pi}{d\theta\; \cos\left( \ell\theta-\frac{\pi}{4}\right) \;\phi(r\cos\theta,r\sin\theta)},\\
	&\pi_\ell(r)=\sqrt{\dfrac{r}{\pi}}\int_0^{2\pi}{d\theta\; \cos\left( \ell\theta-\frac{\pi}{4}\right)\;\pi(r\cos\theta,r\sin\theta)}.
	\end{align}
	The new modes defined this way still satisfy bosonic canonical commutation relations (in other words, the transformation is canonical):
	\begin{equation}
	[\phi_\ell(r),\pi_m(r')]=i\delta(r-r')\delta_{\ell m},\qquad[\phi_\ell(r),\phi_m(r')]=[\pi_\ell(r),\pi_m(r')]=0.
	\end{equation}
	To compute the two-point functions of these new degrees of freedom, we use the following property, which can be proved in a straightforward way. Given any function $h(\zeta)$:
	\begin{equation}
	\int_0^{2\pi}{\int_0^{2\pi}{d\theta\, d\theta'\;\cos\left( \ell\theta-\frac{\pi}{4}\right)\cos\left( \ell'\theta'-\frac{\pi}{4}\right)h(|\theta-\theta'|)}}=\pi\delta_{\ell,\ell'}\int_0^{2\pi}{d\zeta\;h(|\zeta|)\cos{\ell\zeta} }.
	\end{equation}
	Hence we obtain
	\begin{equation}
	\la\phi_\ell(r)\phi_{\ell'}(r')\ra=\delta_{\ell\ell'}\left( \dfrac{\delta(r-r')}{2{\Lambda}}+\sqrt{rr'}\int_{0}^{2\pi}{d\zeta\;f\left( \sqrt{r^2+r'^2-2rr'\cos\zeta}\right) \cos \ell\zeta}\right).
	\end{equation}
	We see that, even though $\phi_\ell$ is not exactly a mode of definite angular momentum, it still only couples to fields with the same value of $\ell$. Something equivalent happens if we include $\pi_\ell$:
	\begin{align}
	&\la\phi_{\ell}(r)\pi_{\ell'}(r')\ra=i\dfrac{\delta(r-r')\delta_{\ell\ell'}}{2}=\overline{\la\pi_{\ell}(r)\phi_{\ell'}(r')\ra},\\
	&\la\pi_{\ell}(r)\pi_{\ell'}(r')\ra=\delta_{\ell\ell'}\left( \dfrac{\delta(r-r')}{2}{\Lambda}+\sqrt{rr'}\int_{0}^{2\pi}{d\zeta\;g\left( \sqrt{r^2+r'^2-2rr'\cos\zeta}\right) \cos l\zeta}\right).
	\end{align}
	Hence the correlation matrix is block diagonal:
	\begin{equation}
	C^{\mc O\mc O'}_{\ell,\ell'}(r,r'):=\la\mc O_\ell(r)\mc O'_{\ell'}(r')\ra=\bigoplus_{\ell=-\infty}^\infty{\la\mc O_\ell(r)\mc O'_\ell(r')\ra}\qquad\mc O,\mc O'\in\{\phi,\pi\}.
	\label{directsum}
	\end{equation}
	When computing entanglement entropies we will make the assumption that the contributions of higher values of $|l|$ decay rapidly in magnitude, so that we can truncate the direct sum above at a small value of $|l|$ \cite{Srednicki1993}.
	
	We will sample the radial indices of the correlation matrix for some lattice spacing $a$ in the same fashion as for 1 dimensional systems (here $\Delta_{\mc O}$ represents the mass dimension of the operator $\mc O$):
	\begin{align}
	C^{\mc O\mc O'}_{\ell,\ell'}(r,r')&\longrightarrow[C^{\mc O\mc O'}_{\ell,\ell'}]_{ij}=a^{\Delta_{\mc O}+\Delta_{\mc O'}}C^{\mc O\mc O'}_{\ell,\ell'}(ia,ja),\\ \delta(r-r')&\longrightarrow\dfrac{\delta_{ij}}{a}\qquad i,j\in\N.
	\end{align}
	We can then start computing contributions from the terms in the direct sum of \eqref{directsum}, by the same symplectic diagonalization procedure explained in Appendix \ref{apa}. We note that indeed the combined sum of all the contributions to $S(x)$ from $|\ell|\leq \ell_\text{max}$ converges as $\ell_\text{max}$ grows, and it does so earlier for smaller values of $x$.
	\subsection{Fermionic theories}
	In the case of fermions, we must take into account that their angular momentum has both an orbital and a spin component. Hence, the irreducible representations of the rotation group $\text{SO}(2)$ will be indexed according to the total angular momentum quantum number $j$. Define
	\begin{align}
	\psi_{1,j}(r)=\sqrt{\dfrac{r}{2\pi}}\int_0^{2\pi}{d\theta\; e^{i\left(j+\frac{1}{2}\right) \theta}\psi_1(r\cos\theta,r\sin\theta)},\\
	\psi_{2,j}(r)=\sqrt{\dfrac{r}{2\pi}}\int_0^{2\pi}{d\theta\;e^{i\left(j-\frac{1}{2}\right) \theta}\psi_2(r\cos\theta,r\sin\theta)},
	\end{align}
	for $j\in \Z+\frac{1}{2}$. This change of variables is unitary, thus it preserves the canonical anticommutation relations:
	\begin{equation}
	\{\psi^{\phantom{\dagger}}_{i,j}(r),\psi^\dagger_{i',j'}(r')\}=\delta_{i,i'}\delta_{j,j'}\delta(r-r'),~~~~~~\{\psi_{i,j}(r),\psi_{i',j'}(r')\}=\{\psi^\dagger_{i,j}(r),\psi_{i',j'}^\dagger(r')\}=0.
	\end{equation}
	$\psi_{i,j}(r)$ are thus fermionic modes with definite total angular momentum. The correlation functions in these new variables read:
	\begin{align}
	&\la\psi^\dagger_{1,j}(r)\psi_{1,j'}^{\prime\phantom{\dagger}}(r')\ra=\sqrt{rr'}\delta_{j,j'}\int_{0}^{2\pi}{d\theta\: e^{-i\left(j+\frac{1}{2}\right) \theta}F\left(\sqrt{r^2+r'^2-2rr'\cos\theta}\right)}\\
	&\la\psi^\dagger_{1,j}(r)\psi_{2,j'}^{\prime\phantom{\dagger}}(r')\ra=\sqrt{rr'}\delta_{j,j'}\int_{0}^{2\pi}{d\theta\: e^{i\left(\xi(r,r',\theta)-j \theta\right) }G\left(\sqrt{r^2+r'^2-2rr'\cos\theta}\right)}\\
	&\la\psi^\dagger_{2,j}(r)\psi_{2,j'}^{\prime\phantom{\dagger}}(r')\ra=\delta(r-r')\delta_{j,j'}-\la\psi^\dagger_{1,j-1}(r)\psi_{1,j'-1}^{\prime\phantom{\dagger}}(r')\ra
	\end{align}
	where \begin{equation}
	\xi(r,r',\theta)=\begin{cases}
	\pi-\dfrac{\theta}{2}-\arctan\dfrac{r\sin\theta}{r'-r\cos\theta},&~~~~ r<r',\vspace{3mm}\\
	\dfrac{\theta}{2}+\arctan\dfrac{r\sin\theta}{r-r'\cos\theta},&~~~~ r>r'.
	\end{cases}
	\end{equation}
	Thanks to rotational symmetry, the two-point functions between modes of  different angular momentum vanish, and thus the correlation matrix again decomposes as a direct sum over different values of $j$:
	\begin{equation}
	C^{ab}_{j,j'}(r,r'):=\la\mc \psi^\dagger_{a,j}(r)\psi^{\phantom{\dagger}}_{b,j'}(r')\ra=\bigoplus_{j\in\Z+\frac{1}{2}}{\la\mc \psi^\dagger_{a,j}(r)\psi^{\phantom{\dagger}}_{b,j}(r')\ra}\qquad a,b=1,2.
	\label{directsumfer}
	\end{equation}
	Now we discretize in the radial variable as done for the bosons. Our expectation is that, for a fixed radius, higher $|j|$ modes will contribute less and less to the entanglement, thus leading to convergence in the entanglement entropy of the disc. This is confirmed by our results displayed in Figure \ref{ent2dfermions}.
	\section{Analytic approximation of entropy scaling at short distances}
	\label{apb}
	In the main text we have stated that expressions can be derived that approximate well the scaling of entanglement entropy in cMERA states for spatial regions $\mc R$ of small sizes compared to the cutoff $1/\Lambda$. Here we present in more detail how this can be achieved. We make use of the techniques reviewed in Appendix \ref{apa}.
	\subsection{Bosons}
	Our strategy consists in getting a reasonably good analytical approximation to the operator $C^{\phi\phi}C^{\pi\pi}$ from Eq. \eqref{opent}, considering it as an operator on $L^2(\mc R)$ with integral kernel:
	\begin{equation}
	K(\vec y,\vec z)=\int_{\mc R}{d\vec w\;C^{\phi\phi}(\vec y,\vec w)C^{\pi\pi}(\vec w,\vec z)}.
	\end{equation}
	\subsubsection{1+1 dimensions}
	Let $\mc R$ be an interval of length $x$ (without loss of generality we consider $\mc R=[0,x]$). For $\Lambda x\ll1$, we consider the eigenvalues of the operator of integral kernel
	\begin{align}
	K(y,z)&=\int_0^x{dy\;C^{\phi\phi}(y,w)C^{\pi\pi}(w,z)}=\nonumber\\&=\int_0^x{dy \left( \dfrac{\delta(y-w)}{2\Lambda}+f_\varepsilon(y-w)\right) \left( \dfrac{\Lambda\delta(w-z)}{2}+g_\varepsilon(w-z)\right)}.
	\end{align}
	Since $|y-w|,|z-w|<x\ll1/\Lambda$, the functions $f_\varepsilon(y-w)$ and $g_\varepsilon(w-z)$ can be well approximated by constants (remember the shape of the correlators in Figure \ref{corr1dbosonsff} and \ref{corr1dbosonspp}), so we write
	\begin{equation}
	f(y,w)\approx A,~~~~g(w,z)\approx B\Lambda^2,
	\end{equation}
	with $A,B$ dimensionless constants. In practice we will Taylor expand $f,g$ around the origin to zeroth order, meaning $A=f_\varepsilon(0),B=g_\varepsilon(0)/\Lambda^2$. The kernel $K(y,z)$ then becomes
	\begin{equation}
	K(y,z)\approx \dfrac{\delta(y-z)}{4}+\dfrac{(A+B)\Lambda}{2}+AB\Lambda^2 x.
	\end{equation}
	To look for the eigenvalues of this approximate kernel over a space of square integrable functions $h(z)\in L^2([0,x])$, we write
	\begin{align}
	\int_0^x{dz\left( \dfrac{\delta(y-z)}{4}+\dfrac{(A+B)\Lambda}{2}+AB\Lambda^2L\right) h(z)}=\lambda h(y)\nonumber\implies\\\implies \left(\dfrac{(A+B)\Lambda}{2}+AB\Lambda^2 x\right)\int_0^x{dz\:h(z)}=\left(\lambda-\dfrac{1}{4} \right) h(y).
	\end{align}
	Now the left hand side does not depend on $y$, so the right hand side must vanish unless $h(y)$ is constant. Thus we find infinitely many eigenvectors\footnote{Namely all $L^2([0,x])$ functions whose integral vanishes.} with eigenvalue $1/4$ and one extra eigenvector (the constant function) with eigenvalue $1/4+(A+B)\Lambda x/2+AB(\Lambda x)^2=(1+2A\Lambda x)(1+2B\Lambda x)/4$. This is the only one that contributes to the entropy, since by Eq. \eqref{eigen}:
	\begin{equation}
	\lambda_i=\dfrac{1}{4}\implies \zeta_i=0\implies \rho_i=\ket{0}\hspace{-0.05mm}\bra{0}\implies S(\rho_i)=0.
	\end{equation}
	Thus we approximate the entropy of the interval by
	\begin{equation}
	S(x)\approx S\left(\lambda=\dfrac{(1+2A\Lambda x)(1+2B\Lambda x)}{4}\right), 
	\end{equation}
	which turns out to be a very good approximation to the entropy at small length scales, as can be seen in Figure \ref{ent1dbosons}. More accurate approximations can be found, for example, by going to higher order in the Taylor expansion of the kernel around $\Lambda x=0$, something that we can do precisely because the correlators in the cMERA state are well-behaved in this limit. 
	\subsubsection{2+1 dimensions}
	We obtain the estimate by approximating the correlators as in the 1+1-dimensional case, and restricting ourselves to the modes with zero angular momentum, $l=0$ (see Appendix \ref{apc}). In fact, the contributions of nonzero values of $l$ vanish for the zeroth order term in the Taylor expansion of the correlators around $|\vec y-\vec z|=0$. Again we approximate the functions $f,g$ in the correlators (Eq. \eqref{f2b} and \eqref{g2b}) by constants
	\begin{equation}
	f(|\vec y-\vec w|)\approx f(0)=:A\Lambda, ~~~~ g(|\vec w-\vec z|)\approx g(0)=:B\Lambda^3, 
	\end{equation}
	for $|\vec y-\vec w|,|\vec w-\vec z|\ll 1/\Lambda$. If we consider $\mc R$ to be a disc centered at the origin with radius $x$, we can change to polar coordinates as in Appendix \ref{apc}. After selecting to look only at modes of zero angular momentum, we have the following approximate operator kernel:
	\begin{align}
	K(r,r'')&=\int_0^x{dr'\:C^{\phi\phi}_{l=l'=0}(r,r')C_{l'=l''=0}^{\pi\pi\phantom{\phi}}(r',r'')}\nonumber\\&\approx \dfrac{\delta(r-r'')}{4}+\left( \pi(A+B)\Lambda^2+2\pi^2 AB\Lambda^4 R^2\right) \sqrt{rr''}
	\end{align}
	defined on a space of square integrable radial functions $h(r)\in L^2([0,x])$. If we look for the eigenvalues of this approximate kernel over this space, we get
	\begin{align}
	\int_0^x{dr''\left[ \dfrac{\delta(r-r'')}{4}+\left( \pi\Lambda^2(A+B)+2\pi^2 AB\Lambda^4 R^2\right) \sqrt{rr''}\right]  h(r'')}=\lambda h(r)\nonumber\implies\\\implies \left( \pi\Lambda^2(A+B)+2\pi^2 AB\Lambda^4 R^2\right)\sqrt{r}\int_0^x{dr''\:\sqrt{r''}h(r'')}=\left(\lambda-\dfrac{1}{4} \right) h(r).
	\end{align}
	Again as in the one dimensional case we find infinitely many eigenvectors with eigenvalue $1/4$ and one extra eigenvector (which in this case is proportional to $\sqrt{r}$) with eigenvalue $1/4+\pi(A+B)(\Lambda x)^2/2+\pi^2 AB(\Lambda x)^4=(1+2\pi A(\Lambda x)^2)(1+2\pi B(\Lambda x)^2)/4$, which is the only one that contributes to the entropy:
	\begin{equation}
	S(x)\approx S\left(\lambda=\dfrac{(1+2\pi A(\Lambda x)^2)(1+2\pi B(\Lambda x)^2)}{4}\right).
	\end{equation}
	In the regime $(\Lambda x\ll1)$, this expression provides a good approximation to the entropy scaling, as can be seen in Figure \ref{fig:ent2dbosons}.
	\subsection{Fermions}
	In the fermionic case, we saw in Appendix \ref{apa} that the entropy can be computed as a sum of contributions coming from the spectrum of the correlation matrix
	\begin{equation}
	C^{ij}(\vec y,\vec z)=\la\psi^\dagger_i(\vec y)\psi^{\phantom{\dagger}}_j(\vec z)\ra\big|^{i,j=1,2}_{x,y\in\mc R}.
	\end{equation}
	This we will now see as the kernel of an integral operator over the space $[L^2(\mc R)]^2$ of pairs of square integrable functions $(h_1(\vec z),h_2(\vec z))$, one per spinor component. 
	\subsubsection{1+1 dimensions}
	Let $\mc R=[0,x]$ with $\Lambda x\ll1$ again and consider the correlation functions at this length scale. The two-point functions can be well approximated by their Taylor expansion to first order around the origin:
	\begin{align}
	\la\psi^\dagger_1(y)\psi_1^{\phantom{\dagger}}(z)\ra&\approx A\Lambda \\\la\psi^\dagger_1(y)\psi^{\phantom{\dagger}}_2(z)\ra&\approx iB\Lambda^2(x-y),
	\end{align} 
	with $A,B\in \R$ dimensionless constants. The linear term is zero for the first one, as is the constant term for the second. The eigenvalue equation for the correlation matrix can then be written as
	\begin{equation}
	\int_0^x{C^{ij}(y,z)h_j(z)\,dz}=\lambda h_i(y),
	\end{equation}
	resulting in
	\begin{align}
	A\lambda\int_0^x{h_1(z)\,dz}+iB\lambda^2\int_0^x{(y-z)h_2(z)\,dz}&=\lambda h_1(x)\\
	 -iB\lambda^2\int_0^x{(y-z)h_1(z)\,dz}-A\lambda\int_0^x{h_2(z)\,dz}&=(\lambda-1) h_2(x).
	\end{align}
	If $\lambda\neq 0,1$ (values with don't contribute to the entropy), these equations constrain the eigenvector to be made of linear functions:
	\begin{equation}
	h_1(z)=a+bz,\qquad h_2(z)=c+dz.
	\end{equation}
	And upon substitution in the equations above we obtain an eigenvalue problem for a 4-dimensional matrix whose solutions are, in a Taylor expansion to first non-vanishing order around $\Lambda x=0$:
	\begin{align}
	\lambda_1&\approx A\Lambda x, ~~~~~~~~~~~~~~\lambda_2\approx 1-A\Lambda x,\nonumber\\
\lambda_3	&\approx -\dfrac{B^2(\Lambda x)^4}{12}, ~~~~~~\lambda_4\approx 1+\dfrac{B^2(\Lambda x)^4}{12}.
	\end{align}
	The two first eigenvalues are in the right interval $[0,1]$ and give the only nontrivial contribution to the entropy. The other two eigenvalues are outside of the acceptable range, but they are so due to high order contributions of $\Lambda x$, what leads us to assume that they are artifacts of the truncation in the correlators and will converge to 0 and 1 respectively if we take more terms in the expansion. Thus our short-range estimation of the entanglement entropy scaling (which ends up being independent of $B$) is 
	\begin{equation}
	S(x)\approx S(\lambda= A\Lambda x)+S(\lambda= 1-A\Lambda x)=2 S(\lambda= A\Lambda x)=2 \dfrac{(1 -\log{A\Lambda x})}{\log{2}}A\Lambda x.
	\end{equation}
	As for its bosonic counterpart, were it needed we could improve on this estimate by using more terms of the Taylor expansions of the correlators.
	\subsubsection{2+1 dimensions}
	To keep the computations simple, and given the results in the 1+1 dimensional case, we approximate the functions $f$ and $g$ from the two-point functions by their zeroth-order Taylor expansion:
		\begin{equation}
		f(|\vec x - \vec y|)\approx f(0)=A\Lambda^2,\qquad g(|\vec x - \vec y|)\approx g(0)=0.
		\end{equation}
	This approximation already implies that the only non-vanishing contributions to the entropy are going to come from the smallest values of angular momentum, namely $j=\pm\frac{1}{2}$ (see Appendix \ref{apb}). In particular, they come from the modes of zero \textit{orbital} angular momentum. The eigenvalue problems that we have to solve are those of the approximate kernels:
	\begin{align}
	\la\psi^\dagger_{1,-\frac{1}{2}}(r)\psi^{\phantom{\dagger}}_{1,-\frac{1}{2}}(r')\ra&\approx 2\pi A\sqrt{rr'},\\
	\la\psi^\dagger_{2,\frac{1}{2}}(r)\psi^{\phantom{\dagger}}_{2,\frac{1}{2}}(r')\ra&\approx\delta(r-r')-2\pi A\sqrt{rr'}.
	\end{align}
	So proceeding as for the bosons we get
	\begin{align}
	S(x)&\approx S(\lambda=\pi A (\Lambda x)^2)+S(\lambda=1-\pi A (\Lambda x)^2)=\nonumber\\&=-2\left(\pi A (\Lambda x)^2\log_2{(\pi A (\Lambda x)^2)}+(1-\pi A (\Lambda x)^2)\log_2{(1-\pi A (\Lambda x)^2)} \right) 
	\end{align}
	which provides a good estimation in the short distance regime (Figure \ref{ent2dfermions}).

	Notice that in both the bosonic and fermionic case we have approximated the entanglement entropy for that of a theory with flat two point functions. Because we assume such simple situation, the entropy can be computed easily, and we end up with expressions that depend in the same functional way on the volume of the spatial region ($\Lambda x$ in one spatial dimension, $\pi(\Lambda x)^2$ in two spatial dimensions) for each statistics (bosons or fermions).
	
	\acknowledgments
	We thank Qi Hu, Nick van den Broeck and Michael Lamoureux for enlightening conversations. We also thank the JHEP referee for many useful comments, particularly for suggesting the use of asymptotic expansions. A. Franco-Rubio is supported by the La Caixa Graduate Fellowship Program. The authors also acknowledge support by the Simons Foundation (Many Electron Collaboration) and Compute Canada. Research at Perimeter Institute is supported by the Government of Canada through Industry Canada and by the Province of Ontario through the	Ministry of Research and Innovation.
	
	\bibliographystyle{JHEP}	
	\bibliography{articulouno}

\end{document}